\documentclass[12pt,preprint]{aastex}
\usepackage{graphicx,subfigure,amsmath,natbib}
\newcommand\gsim{\,\lower3pt\hbox{$\sim$}\llap{\raise2pt\hbox{$>$}}\,}
\newcommand\lsim{\,\lower3pt\hbox{$\sim$}\llap{\raise2pt\hbox{$<$}}\,}

\begin{document}
\title{Three-Dimensional MHD Simulations of Emerging Active Region Flux in a
Turbulent Rotating Solar Convective Envelope: the Numerical Model and Initial
Results}

\author{Y. Fan, N. Featherstone\altaffilmark{2}, and F. Fang}
\affil{High Altitude Observatory, National Center for Atmospheric
Research\altaffilmark{1}, 3080 Center Green Drive, Boulder, CO 80301}

\altaffiltext{1} {The National Center for Atmospheric Research
is sponsored by the National Science Foundation}
\altaffiltext{2} {Now at JILA, University of Colorado at Boulder}

\begin{abstract}
We describe a 3D finite-difference spherical anelastic MHD (FSAM) code
for modeling the subsonic dynamic processes in the solar convective envelope.
A comparison of this code with the widely used global spectral anlastic
MHD code, ASH (Anelastic Spherical Harmonics), shows that FSAM produces
convective flows with statistical properties and mean flows similar to
the ASH results.
Using FSAM, we first simulate the rotating solar convection in a
partial spherical shell domain and obtain a statistically steady,
giant-cell convective flow with a solar-like differential rotation.
We then insert buoyant toroidal flux tubes
near the bottom of the convecting envelope and simulate the rise of
the flux tubes in the presence of the giant cell convection.
We find that for buoyant flux tubes with an initial field strength of
100 kG, the magnetic buoyancy largely determines the rise of the
tubes although strong down flows produce significant
undulation and distortion to the shape of the emerging $\Omega$-shaped loops.
The convective flows significantly reduce the rise time it takes for the apex
of the flux tube to reach the top.
For the weakly twisted and untwisted cases we simulated, the apex portion is
found to rise nearly radially to the top in about a month, and produce an
emerging region (at a depth of about 30 Mm below the photosphere) with an
overall tilt angle consistent with the active region
tilts, although the emergence pattern is more complex compared to the case
without convection. Near the top boundary at a depth of about 30 Mm, the
emerging flux shows a retrograde zonal flow of about 345 m/s relative to
the mean flow at that latitude.

\end{abstract}

\section{Introduction}

If we believe that active regions on the solar surface originate from a strong
toroidal magnetic field generated at the base of the convection zone by the
solar dynamo mechanism, then we need to understand how active-region-scale flux
tubes rise through the turbulent solar convection zone to the surface. Recently
significant insight has been gained in this area from a series of work
\citep{Weber:Fan:Miesch:2011, Weber:Fan:Miesch:2012}
conducted using a thin flux tube model driven via the drag force term
by a time dependent giant-cell convective flow with a solar like differential
rotation, computed separately from a 3D global convection simulation
with the Anelastic Spherical Harmonic (ASH) code \citep{Miesch:etal:2006}.
Because of the low computational cost for the 1-D thin flux
tube model, a large number of simulations of rising flux tubes with a range
of initial field strengths, fluxes, initial latitudes, and sampling different
time spans of the convective flow field were carried out. Meaningful
statistics on the properties of the emerging tubes in regard to the latitude
of emergence, tilt angles, apparent zonal motion, and clustering in longitudes
of emergence (i.e. active longitudes) is obtained.
It is found that the dynamic evolution of the flux tube changes from convection dominated to magnetic buoyancy dominated as the initial field strength
increases from 15 kG to 100 kG. At 100 kG, the development of $\Omega$-shaped
rising loops is mainly controlled by the growth of the magnetic buoyancy
instability, with the strongest convective downdrafts capable of producing
moderate undulations on the emerging loops.
It is found that although helical convection promotes mean tilts towards
the observed Joy's law trend, results still favor stronger fields ($> 40$ kG)
for the initial toroidal tubes to avoid too large a tilt
angle scatter produced by convection to be consistent with the observations
\citep{Weber:Fan:Miesch:2012}.

Although the thin flux tube model essentially preserves the frozen-in condition
for the evolution of the flux tube and allows for a large number of simulations
to achieve meaningful statistics, it is highly idealized. It ignores the 3D
nature of the magnetic field evolution and assumes the tube is a cohesive
object.  In parallel to the thin flux tube calculations, several
self-consistent 3D global MHD simulations of rising flux
tubes in a rotating spherical shell of solar convection and the associated mean
flows have been carried out \citep{Jouve:Brun:2009, Jouve:etal:2013} using
the ASH code.
These simulations study rather large flux tubes (significantly greater than
the flux contained in typical active regions) due to the limited numerical
resolutions in the global scale simulations.
It is found that the rise velocity and the characteristics of the emerging
loops are strongly affected by the convective motions when loops of less than
$10^5$ G are considered.
In addition, the question of how strong buoyant flux tubes can
self-consistently form from dynamo generated mean field and rise to the
surface is also being addressed in a set of full global convective dynamo
simulations of a fast rotating stellar envelope with 3 times the solar
rotation rate
\citep{Nelson:etal:2011, Nelson:etal:2013, Nelson:etal:solphys:2013},
also using the ASH code.

In this paper, we describe a new 3D Finite-difference Spherical
Anelastic MHD (FSAM) code for modeling the subsonic dynamics of 
the turbulent solar convective envelope. The code uses a modified
Lax-Friedrichs scheme (as described in the Appendix) for computing the
upwinded fluxes in the advection terms, which allows for stable numerical
integration of the anelastic MHD equation with no explicit diffusion.
Of course numerical diffusion is present, but is minimized in smooth
regions which helps to preserve the frozen-in condition of the magnetic
field evolution in the simulations of rising flux tubes.
We carry out a comparison between the FSAM code and the ASH code with
a simulation of rotating convective flow in a spherical shell. We find that
even with the absence of the polar region (necessary due to the polar singularity
associated with the latitude-longitude grid discretization), the FSAM code
can produce convective flows with similar statistical properties and mean
flow properties as the fully global ASH spectral code.
We then use the FSAM code to conduct a simulation of rotating solar
convection in a spherical shell wedge domain driven at the lower boundary 
by the diffusive heat flux corresponding to the solar luminosity. We obtain
a statistically steady solution of giant-cell convection with a solar-like
differential rotation. Into the giant-cell convective flow, we then insert
buoyant toroidal flux tubes with an initial field strength of $10^5$ G near
the bottom of the envelope, to study how the tubes rise under the
presence of convection.  We find that the buoyant loops rise based
on the initial magnetic buoyancy distribution and also are significantly
reshaped by the strong convective downdrafts. They can rise to the surface
nearly radially, and produce emerging regions with radial flux distribution
of the two polarities that are consistent with the observed mean tilt angles
of solar active regions.
At a depth of about 30 Mm below the photosphere, the emerging flux shows
a retrograde zonal motion in the midst of the prograde flow of the banana
cells, with a speed of $\sim - 350$ m/s relative to the mean plasma
zonal flow at the emerging latitude.

\section{The Numerical Model\label{sec:model}}
We solve the following anelastic MHD equation in a spherical shell domain:
\begin{equation}
\nabla \cdot ( \rho_0 {\bf v} ) = 0,
\label{eq:continuity}
\end{equation}
\begin{equation}
\rho_0 \left [ \frac{\partial {\bf v}}{\partial t} + ({\bf v} \cdot \nabla )
{\bf v} \right ]
= 2 \rho_0 {\bf v} \times {\bf \Omega} - \nabla p_1 + \rho_1 {\bf g}
+ \frac{1}{4 \pi} ( \nabla \times {\bf B}) \times {\bf B}
+ \nabla \cdot {\cal D}
\label{eq:momentum}
\end{equation}
\begin{equation}
\rho_{0} T_0 \left [ \frac{\partial s_1}{\partial t}
 + ( {\bf v} \cdot \nabla ) (s_0 + s_1 ) \right ]
= \nabla \cdot ( K \rho_0 T_0 \nabla s_1 ) - ( {\cal D} \cdot \nabla )
\cdot {\bf v} + \frac{1}{4 \pi} \eta ( \nabla \times {\bf B} )^2
- \nabla \cdot {\bf F}_{\rm rad}
\label{eq:entropy1}
\end{equation}
\begin{equation}
\nabla \cdot {\bf B} = 0
\label{eq:divB}
\end{equation}
\begin{equation}
\frac{\partial {\bf B}}{\partial t} = \nabla \times ( {\bf v} \times {\bf B} )
- \nabla \times ( \eta \nabla \times {\bf B} ) ,
\label{eq:induction1}
\end{equation}
\begin{equation}
\frac{\rho_1}{\rho_0} = \frac{p_1}{p_0} - \frac{T_1}{T_0},
\label{eq:eqstate}
\end{equation}
\begin{equation}
\frac{s_1}{c_p} = \frac{T_1}{T_0} - \frac{\gamma -1}{\gamma}
\frac{p_1}{p_0} ,
\label{eq:2ndthermlaw}
\end{equation}
where $s_0 (r)$, $p_0 (r)$, $\rho_ 0 (r)$, $T_0 (r)$, and
${\bf g} = - g_0 (r) {\hat {\bf r}}$ denote
the profiles of entropy, pressure, density, temperature, and the gravitational
acceleration of a time-independent, reference state of hydrostatic equilibrium
and nearly adiabatic stratification, $c_p$ is the specific heat capacity at
constant pressure, $\gamma$ is the ratio of specific
heats, and ${\bf v}$, ${\bf B}$, $s_1$, $p_1$,
$\rho_1$, and $T_1$ are the dependent variables of velocity, magnetic field,
entropy, pressure, density, and temperature to be solved that describe the
changes from the reference state. In equation (\ref{eq:momentum}),
${\bf \Omega}$ denotes the solid body
rotation rate of the Sun and is the rotation rate of the frame of reference,
where $\Omega = 2.7 \times 10^{-6} {\rm rad} \; {\rm s}^{-1}$,
and ${\cal D}$ is the viscous stress
tensor:
\begin{equation}
{\cal D}_{ij} = \rho_0 \nu \left [ S_{ij} - \frac{2}{3} ( \nabla \cdot
{\bf v} ) \delta_{ij} \right ] ,
\label{eq:visstress}
\end{equation}
where $\nu$ is the kinematic viscosity, $\delta_{ij}$ is the unit tensor, and
$S_{ij}$ is given by the following in spherical polar coordinates:
\begin{equation}
S_{rr} = 2 \frac{\partial v_r}{\partial r}
\end{equation}
\begin{equation}
S_{\theta \theta} = \frac{2}{r} \frac{\partial v_{\theta}}{\partial \theta}
+ \frac{2 v_r}{r}
\end{equation}
\begin{equation}
S_{\phi \phi} = \frac{2}{r \sin \theta} \frac{\partial v_{\phi}}{\partial \phi}
+ \frac{2 v_r}{r} + \frac{2 v_{\theta}}{r \sin \theta } \cos \theta
\end{equation}
\begin{equation}
S_{r \theta} = S_{\theta r} = \frac{1}{r} \frac{\partial v_r}{\partial \theta}
+ r \frac{\partial}{\partial r} \left ( \frac{v_{\theta}}{r} \right )
\end{equation}
\begin{equation}
S_{\theta \phi } = S_{\phi \theta} = \frac{1}{r \sin \theta}
\frac{\partial v_2} {\partial \phi } + \frac{\sin \theta} {r}
\frac{\partial}{\partial \theta} \left ( \frac{v_{\phi}}{\sin \theta} \right )
\end{equation}
\begin{equation}
S_{\phi r} = S_{r \phi } = \frac{1}{r \sin \theta}
\frac{\partial v_r}{\partial \phi} + r \frac{\partial}{\partial r}
\left ( \frac{v_{\phi}}{r} \right ) .
\end{equation}
Futhremore, $K$ in equation (\ref{eq:entropy1}) denotes the thermal
diffusivity, and $\eta$ in equations (\ref{eq:induction1}) and
(\ref{eq:entropy1}) denotes the magnetic diffusivity.  The last term in
equation (\ref{eq:entropy1}) is a heating source term due to the radiative
diffusive heat flux ${\bf F_{\rm rad}} $
in the solar interior, where 
\begin{equation}
{\bf F_{\rm rad}} = \frac{16 \sigma_s {T_0}^3}{3 \kappa \rho_0} \nabla T_0 ,
\label{eq:radflux}
\end{equation}
and $\sigma_s$ is the Stephan-Boltzman constatn, $\kappa$
is the Rosseland mean opacity.

Using equations (\ref{eq:eqstate}) and (\ref{eq:2ndthermlaw}) to express
$\rho_1$ in terms of $p_1$ and $s_1$ in equation (\ref{eq:momentum}), and after
some manipulations using the ideal gas law and hydrostatic balance for the
reference state, we obtain
\begin{eqnarray}
\rho_0 \left [ \frac{\partial {\bf v}}{\partial t} + ({\bf v} \cdot \nabla )
{\bf v} \right ]
& = & 2 \rho_0 {\bf v} \times {\bf \Omega}
- \rho_0 \nabla \left ( \frac{p_1}{\rho_0} \right )
+ \rho_0 g_0 \frac{s_1}{c_p} {\hat {\bf r}}
\nonumber \\
& & + \frac{1}{4 \pi} ( \nabla \times {\bf B}) \times {\bf B}
+ \nabla \cdot {\cal D} .
\label{eq:momentumsolv1}
\end{eqnarray}
Note, in deriving the above equation, we have ignored terms of higher order
in $\delta$, where
\begin{equation}
\delta \equiv \frac{{\rm d} \ln T_0}{{\rm d} \ln p_0}
- \frac{\gamma -1}{\gamma}
\end{equation}
is the non-dimensional super-adiabaticity of the reference
stratification, and its magnitude is $ \ll 1$ in the anelatic approximation.
The super-adiabaticity $\delta$
is related to the entropy gradient of the reference state as follows:
\begin{equation}
\frac{{\rm d} s_0}{{\rm d} r} = - c_p \frac{\delta}{H_{p0}},
\end{equation}
where
\begin{equation}
H_{p0} = - \left ( \frac{{\rm d} \ln p_0}{{\rm d} r} \right )^{-1}
\end{equation}
denotes the pressure scale height.

To ensure the divergence free condition of equation (\ref{eq:continuity}) is
satisfied, $p_1$ in equation (\ref{eq:momentumsolv1}) needs to satisfy the
following linear elliptic equation, which we solve at every time step before
using it in the above momentum equation to advance ${\bf v}$:
\begin{equation}
\nabla \cdot \left [ \rho_0 \nabla \left ( \frac{p_1}{\rho_0} \right )
\right ] = \nabla \cdot {\cal F}
\label{eq:pressuresolv}
\end{equation}
where
\begin{equation}
{\cal F} = - \rho_0 {\bf v} \cdot \nabla {\bf v} + 2 \rho_0
{\bf v} \times {\bf \Omega} + \rho_0 g_0 \frac{s_1}{c_p} {\hat {\bf r}}
+ \frac{1}{4 \pi} ( \nabla \times {\bf B}) \times {\bf B}
+ \nabla \cdot {\cal D} .
\label{eq:rhs_pressuresolv}
\end{equation}
Also applying the divergence free condition of
equation (\ref{eq:continuity}), we
can rewrite the entropy equation as follows:
\begin{eqnarray}
\rho_0 T_0 \frac{\partial s_1}{\partial t}
& = & - \nabla \cdot [ \rho_0 {\bf v} T_0 (s_1 + s_0 )] - \rho_0 v_r
(s_1 + s_0 ) \frac{g_0}{c_p}
\nonumber \\
& & + \rho_0 \nu \left [ {S_{r \theta}}^2 + {S_{\theta \phi}}^2
+ { S_{\phi r} }^2 + \frac{1}{6} ( (S_{rr}-S_{\theta \theta} )^2
+ \left ( S_{\theta \theta} - S_{\phi \phi})^2 + (S_{\phi \phi} - S_{rr})^2
\right ) \right ]
\nonumber \\
& & + \eta (\nabla \times {\bf B})^2 + \nabla \cdot ( K \rho_0 T_0 \nabla s_1 )
+ \nabla \cdot \left ( \frac{16 \sigma_s {T_0}^3}{3 \kappa \rho_0}
\nabla T_0 \right ) .
\label{eq:entropy1solv}
\end{eqnarray}
In deriving the above equation, we have used
\begin{equation}
\frac{{\rm d} T_0}{{\rm d} r} = - \frac{g_0}{c_p}
\end{equation}
where we have ignored the terms of order $O(\delta)$ produced by the
small superadiabaticity in the reference profile of $T_0$ and only
preserved the zeroth order term (corresponding to the adiabatic
stratification). The viscous heating term, which is positive definite, has
also been written out explicitly in terms of the tensor components $S_{ij}$.

Thus, in summary, we numerically solve equations
(\ref{eq:momentumsolv1}), (\ref{eq:pressuresolv}), (\ref{eq:entropy1solv}),
and (\ref{eq:induction1}), to advance the dependent variables ${\bf v}$,
$p_1$, $s_1$, and ${\bf B}$.  A more detailed description of the numerical
schemes used to solve these equations is given in the appendix.
We further note that by summing
${\bf v} \cdot$ equation (\ref{eq:momentumsolv1}), ${\bf B} \cdot $ equation
(\ref{eq:induction1}), and equation (\ref{eq:entropy1solv}),
we can also obtain the following equation for total energy conservation:
\begin{eqnarray}
& & {\partial \over \partial t} \left ( \rho_0 \frac{v^2}{2}
+ \rho_0 T_0 s_1 + \frac{B^2}{8 \pi } \right )
\nonumber \\
& = & - \nabla \cdot \left [ \left ( \rho_0 \frac{v^2}{2} + p_1
+  \rho_0 T_0 (s_1+s_0) \right ) {\bf v} 
- \frac{1}{4 \pi} ( {\bf v} \times {\bf B} ) \times {\bf B} \right ]
- \rho_0 v_r s_0 \frac{g_0}{c_p}
\nonumber \\
& & + \nabla \cdot ( {\bf v} \cdot {\cal D} )
+ \nabla \cdot \left [ - \eta ( \nabla \times {\bf B} ) \times {\bf B} \right ]
\nonumber \\
& & + \nabla \cdot \left ( K \rho_0 T_0 \nabla s_1 \right )
+ \nabla \cdot \left ( \frac{16 \sigma_s {T_0}^3}{3 \kappa \rho_0}
\nabla T_0 \right ) .
\label{eq:etot_dim}
\end{eqnarray}
Since numerically we are solving the entropy equation (\ref{eq:entropy1solv})
instead of the above total energy equation explicitly in conservative form,
the total energy equation can serve as an independent check on the
effects of numerical dissipation.

\section{A comparison of FSAM and ASH \label{sec:ashcompare}}
Were FSAM to include the polar caps, we could assess the accuracy of its
results by direct comparison against the hydrodynamic anelastic benchmark
solution of \citet{Jones:etal:2011}.  The benchmark solution is most appropriate
for solution domains encompassing the full sphere as it manifests as a
sectoral mode of convection, localized around the equator, and propagating
prograde with time.  Unfortunately, we find that the absence of a polar region
in FSAM alters the meridional circulations achieved in the benchmark solution,
ultimately preventing FSAM from obtaining the pure spherical harmonic mode
of convection achievable when the full sphere is simulated.

One major use of FSAM is for studies of magnetic flux emergence through a
turbulent solar convection zone, and while benchmarks are of some interest,
we are most concerned with its ability to yield convective motions with
properties similar to those thought to exist in the Sun.  To this end, we
have chosen to run a somewhat more turbulent simulation and
compare the properties of the solution against those obtained using the
Anelastic Spherical Harmonic (ASH) code.  ASH solves the three-dimensional
(3-D) anelastic MHD equations in deep spherical shells using a pseudospectral
approach.  It employs a spherical harmonic expansion in the horizontal
direction, and Chebyshev polynomials or a finite-difference approach in the
radial direction.  ASH has been used extensively to model the solar convection
zone \citep[e.g.][]{Brun:etal:2004, Miesch:etal:2008},
and has shown good agreement
with other anelastic codes when applied to the benchmark problems of
\citet{Jones:etal:2011}. By comparing the results of FSAM against ASH in a
somewhat more turbulent regime than the weakly non-linear benchmark test
in \citet{Jones:etal:2011}, we anticipate that the properties of the
convective flows in the bulk of the solution is less affected
by the role played by the polar region.

\subsection{Experimental Setup}
We have constructed a comparison experiment by modeling convection in a
spherical shell spanning the full depth of the solar convection zone, albeit
with a much reduced density stratification relative to the Sun. We assume 
that the gravitational acceleration $g_0(r)$ varies as $GM_{int}/r^2$
within the shell, where $M_{int}$ is the mass interior to the base of the
convection zone and $G$ is the gravitational constant, and use the
following adiabatically stratified, polytropic atmosphere as the reference
state:
\begin{equation}
\label{polytrope}
\rho_0 = \rho_i\left(\frac{\zeta}{\zeta_i}\right)^n,~~~~T_0=T_i\frac{\zeta}{\zeta_i},~~~~p_0=p_i\left(\frac{\zeta}{\zeta_i}\right)^{n+1},
\end{equation}
where the subscript ``i" denotes the value of a quantity at the inner boundary,
and n is the polytropic index.  The radial variation of the reference state
is captured by the variable $\zeta$, defined as
\begin{equation}
\zeta=c_0+\frac{c_1d}{r},
\end{equation}
where $d = r_o-r_i$ is the depth of the convection zone.  The constants $c_0$ and $c_1$ are given by
\begin{equation}
c_0=\frac{2\zeta_o-\beta-1}{1-\beta},~~~~c_1=\frac{(1+\beta)(1-\zeta_o)}{(1-\beta)^2},
\end{equation}
with
\begin{equation}
\zeta_o=\frac{\beta+1}{\beta\mathrm{exp}(N_\rho/n)+1},~~~~\zeta_i=\frac{1+\beta-\zeta_o}{\beta}.
\end{equation}
Here $\zeta_i$ and $\zeta_o$ are the values of $\zeta$ on the inner and outer
boundaries, $\beta=r_i/r_o$, and $N_\rho$ is the number of density scale heights
across the shell. Further details of this model setup may be found in
Jones et al. (2011).

We choose to employ this description for the reference state, with a density
variation of one scale height across the shell.  Entropy ($s_1$) is fixed at
both the upper and lower boundary with a constant entropy difference $\Delta S$
across the domain, allowing us to specify a
Rayleigh number.  Our model further differs from the Sun in that radiative
heating by photon diffusion (the last term on the right hand side of
eq. [\ref{eq:entropy1solv}]), particularly important near
the base of the convection zone, is neglected.  The thermal energy throughput
of the system is instead entirely determined by thermal conduction (the
2nd to last term on the right hand side of eq. [\ref{eq:entropy1solv}]) at the
boundaries.  The degree of thermal conduction may vary in time (due to the
changes in $\partial s_1 / \partial r$ at the boundaries), but reaches a
statistically steady state that is itself determined by the entropy gradients
established by the convection.  Our thermal diffusivity $K$ and viscosity
$\nu$ are taken to be constant functions of depth with a Prandtl number
$P_r$ of unity. Values for the simulation parameters are provided in
Table \ref{inputs_table}.

For the ASH simulation, we used 200 points in radius, and chose a maximum
spherical harmonic degree of 170, yielding an effective resolution of
200$\times$256$\times$512 ($r$,$\theta$,$\phi$). To assess the effects of
resolution and polar region removal, we chose to run three distinct FSAM
simulations.  These three simulations were identical in every respect except
for spatial resolution and domain size.  The primary simulation (case A),
employed a resolution that is approximately half that of the ASH simulation,
with a resolution of 96$\times$128$\times$256, and extended to
$\pm$60$^{\circ}$ in latitude.  Case B extended over the same latitude range,
but employed a resolution of 192$\times$256$\times$512 (twice that of case A).
Case C extended from $\pm$75$^{\circ}$ in latitude, and employed a resolution
of 96$\times$160$\times$256 (similar to case A).  With the exception of
case B, each simulation was evolved for 4000 days (about seven thermal and
viscous diffusion times) to ensure that a thermally and dynamically
well-equilibrated state was obtained.  The somewhat more expensive Case B was
evolved for 1200 days (about two thermal diffusion times).

\subsection{Convective Morphology}
\begin{figure}
\centering
\includegraphics[width=\textwidth]{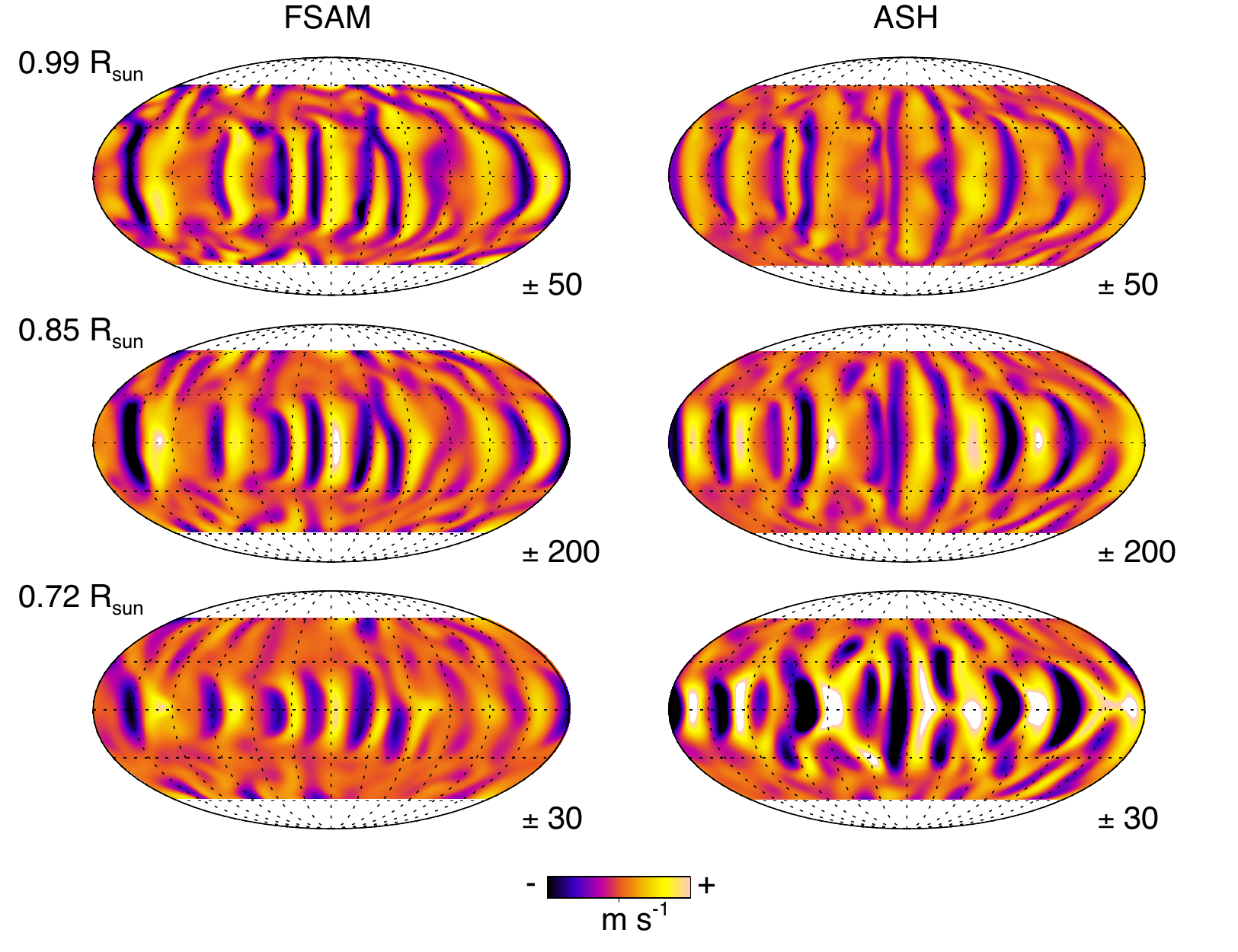}
\caption{Radial velocity $v_{r}$ at three depths from FSAM case A and the ASH simulation, shown in Mollweide projection.  Horizontal lines indicate lines of constant latitude, and curved arcs lines of constant longitude.  Regions of yellow denote upflows, and regions of blue indicate downflows.  Flows in FSAM are stronger near the surface and weaker near the bottom than their counterparts in ASH.  Banana cell patterns in ASH tend to reach to higher latitudes than those in FSAM.
\label{vr_shells}}
\end{figure}

A survey of the radial flows realized in ASH and case A is illustrated in
Figure \ref{vr_shells}.  Here snapshots of radial velocity $v_r$, taken at
three depths from one instant in time near the end of each simulation, are
shown.  We have omitted the polar regions of ASH in this plot for ease of
comparison.  Near-surface flow structures are similar in ASH and FSAM results,
with flows in both simulations achieving amplitudes of roughly 50 m s$^{-1}$
at the top of the domain.  Banana cell-like patterns are prominent at the
equator in each simulation, but extend to somewhat higher latitudes in the
ASH results, possibly due to the inclusion of the polar regions.  At
mid-convection zone, flows are comparable in amplitude, but the disparity in
latitudinal extent of the banana cells has grown, and near the base of the
simulation, the solutions are markedly different.  Radial flows in FSAM at
this depth are both weaker than those realized in ASH and are largely
confined to a much narrower equatorial region.

\begin{figure}
\centering
\includegraphics[width=\textwidth]{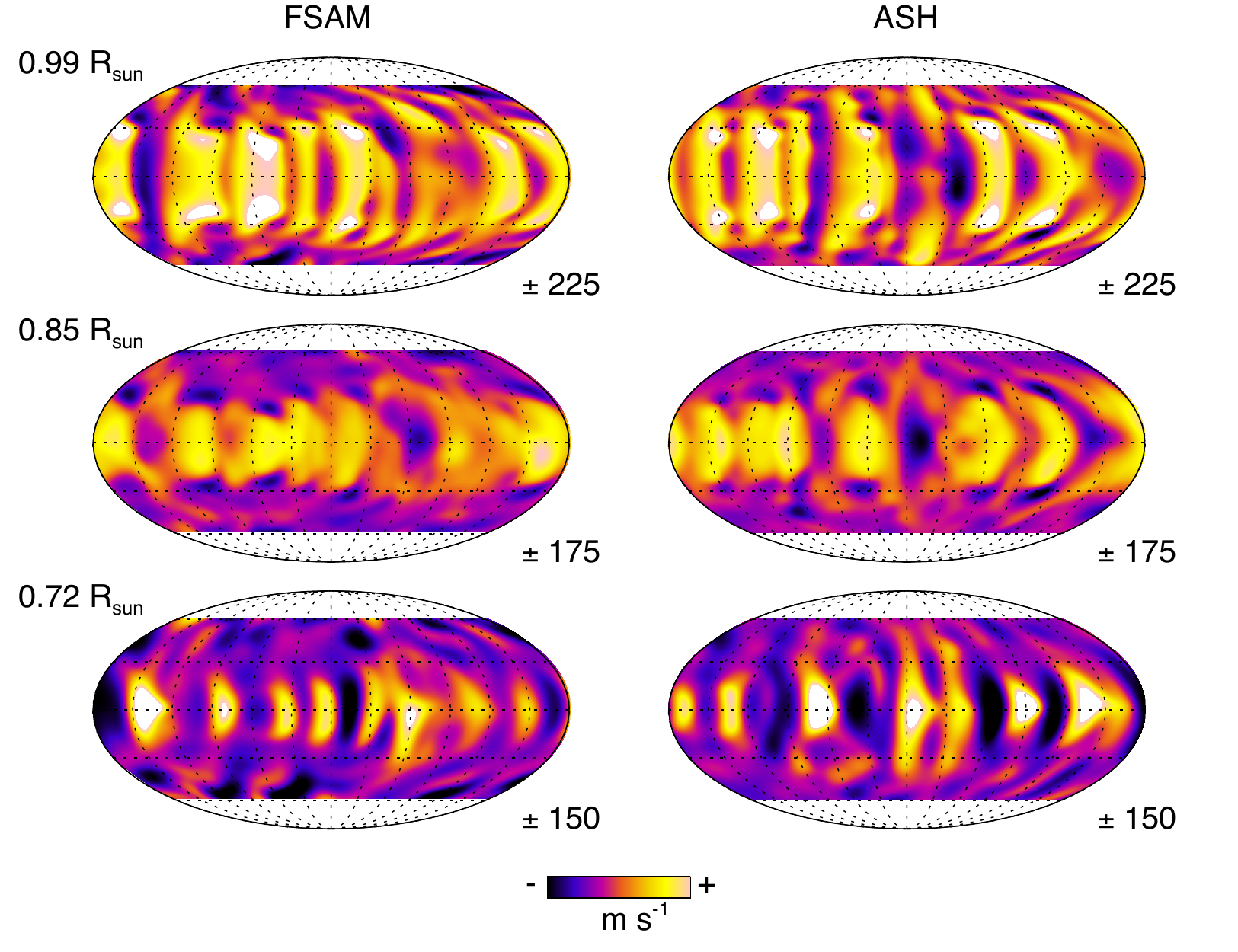}
\caption{Zonal velocity $v_{\phi}$ plotted at three depths in each simulation at the same time instant as Fig. \ref{vr_shells}.  Azimuthal flows in FSAM case A are similar to those of ASH throughout most of the convection zone, becoming somewhat stronger near the surface.
\label{fig:vphi_shells}}
\end{figure}

The zonal velocity $v_{\phi}$ for each simulation is shown in
Figure \ref{fig:vphi_shells}.  Both simulations develop a prograde differential
rotation at the equator.  Flow amplitudes compare at depth in a similar
manner to the radial flows, with the prograde region of differential rotation
occupying a smaller latitudinal extent near the base of the convection zone
in the FSAM results than in the ASH results.  There are two effects that
contribute to the differences in these two solutions, the most obvious of
which is the absence of a polar region in FSAM.  Moreover, the numerical
diffusion scheme employed in FSAM, which operates in addition to the explicit
diffusivities, can lead to differing results where the simulation is
under-resolved.

\begin{figure}
\centering
\includegraphics[width=3.0truein]{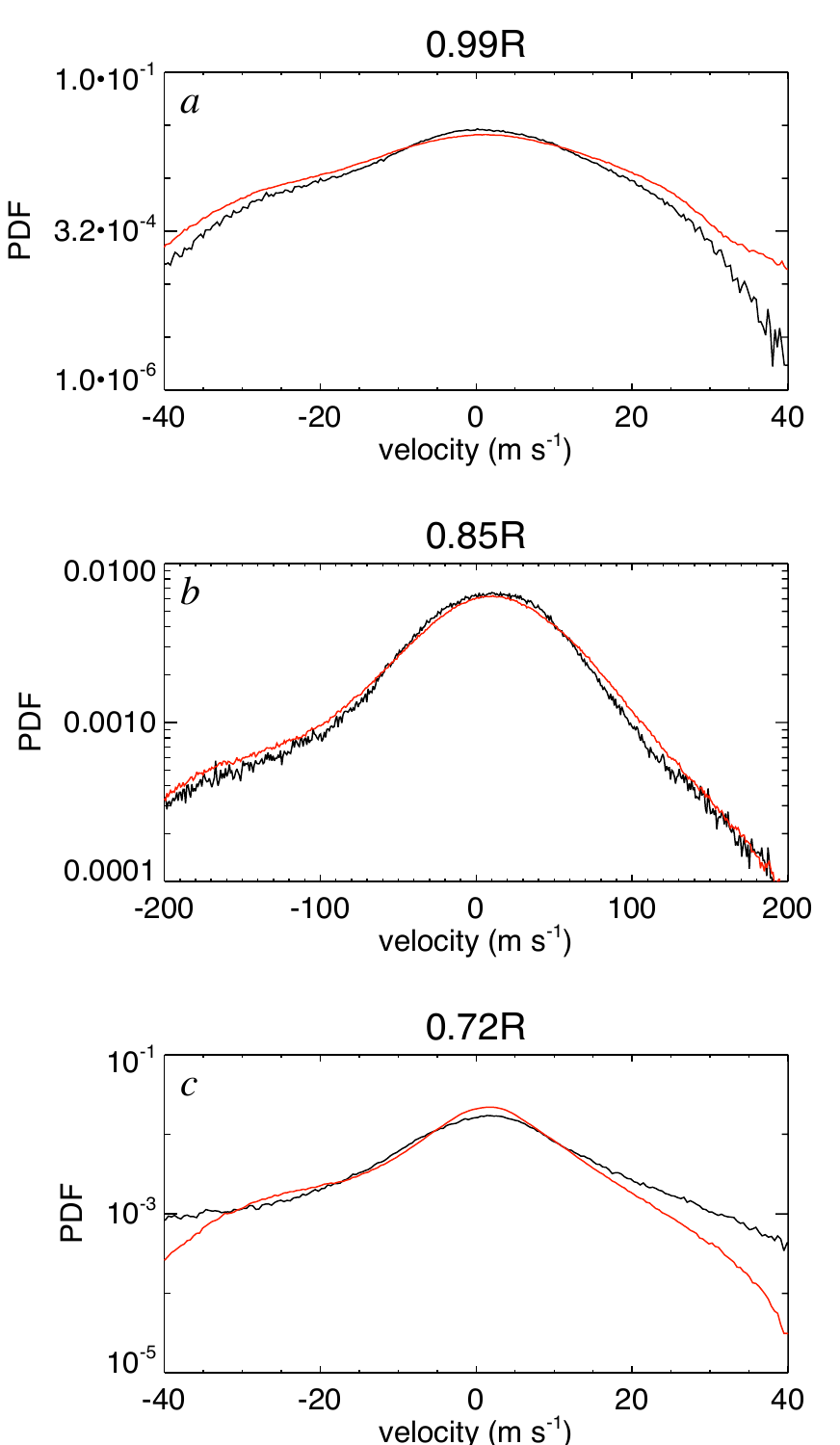}
\caption{Probability distribution functions (PDF) taken near \textit{(a)} the
top, \textit{(b)} middle, and \textit{(c)} bottom of each simulation, averaged
in time over 10 rotation periods.  PDFs for FSAM case A are shown in red.
PDFs for ASH are plotted in black and have been computed over the FSAM range of
latitudes.  Both simulations show good agreement in the mid-convection zone.
The cores of the distributions agree well in the boundary layers, though the
wings are stronger for FSAM in the upper boundary layer and weaker relative
to ASH in the lower boundary layer.
\label{fig:pdfs}}
\end{figure}

These snapshots of the flow are from but one instant in each simulation.
A sense of how the two solutions compare in a statistical sense may be better
gained by looking at probability distribution functions (PDFs) of radial
velocity.  Figure \ref{fig:pdfs} depicts PDFs of $v_r$, averaged over 500 days
of evolution at the end of each simulation, and shown at the same three
depths as Figure \ref{vr_shells}.  Case A (red) exhibits substantially higher
power in the wings of its PDF near the surface than does ASH.
At mid-convection zone, the two distributions are in much better agreement,
although the FSAM simulation still exhibits somewhat stronger wings.
Near the base of the convection zone, where the flows are noticiably different
in Figure \ref{vr_shells}, the ASH flows exhibit significantly stronger wings.

\begin{figure}
\centering
\includegraphics[width=3.0truein]{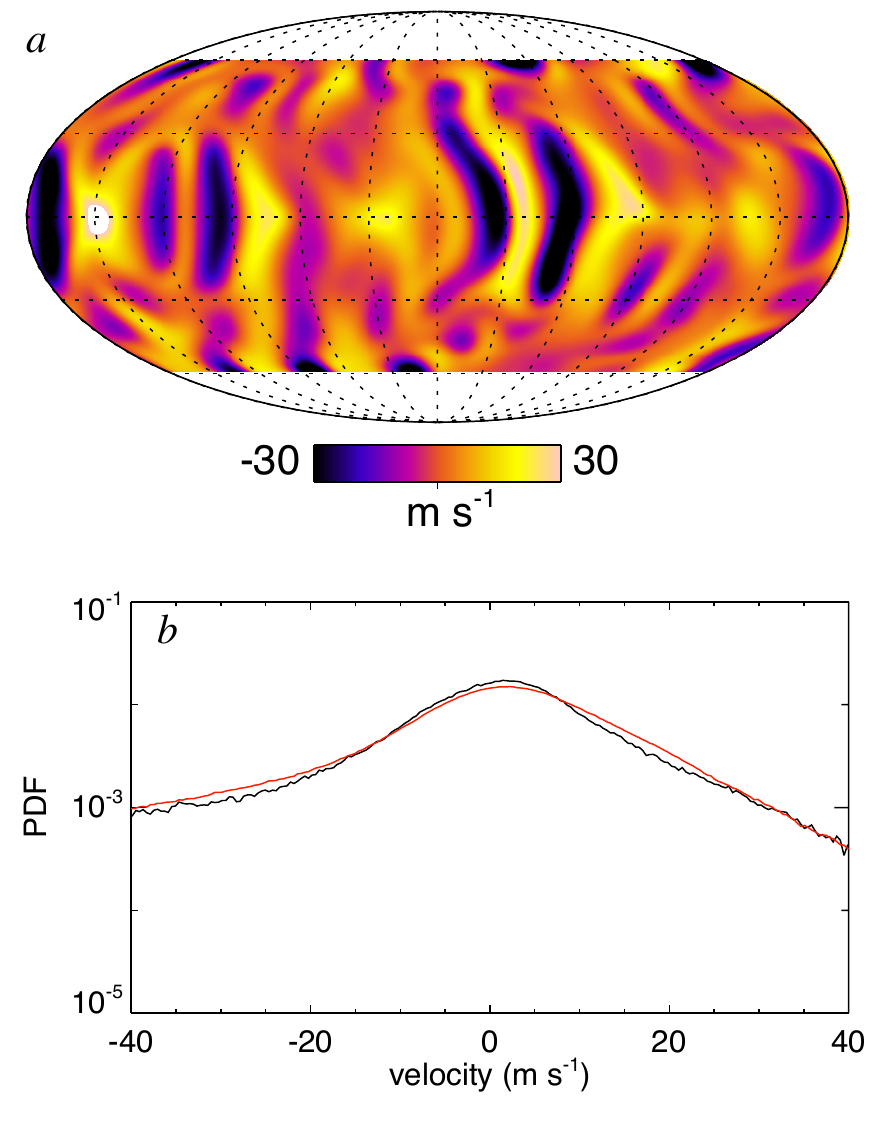}
\caption{The effect of doubling the spatial resolution.  \textit{(a)} Radial velocity $v_r$ from FSAM case B near the bottom of the convection zone (0.72R).  Flows are noticiably stronger than case A and extend to higher latitudes when the spatial resolution is doubled.  \textit{(b)} PDFs of radial velocity (at 0.72R) for FSAM case B (red), and the ASH simulation (black).  The high velocity wings of the distribution have been enhanced substantially with respect to case A, reaching good agreement with those of the ASH simulation.
\label{fig:hres_hist}}
\end{figure}

We find that these differences in the lower convection zone are substantially diminished when the
spatial resolution of the simulation is doubled.  We suspect that flow
structures associated with convective downdrafts impacting the impenetrable
lower boundary are under-resolved in case A, especially in the horizontal
dimensions where the resolution is about 4 times worse than the vertical
resolution.  Figure \ref{fig:hres_hist}$a$ depicts $v_r$ near the base of the
convection zone for FSAM case B, which has twice the spatial resolution of
case A.  Flow amplitudes and structure sizes are comparable to the ASH case
for case B.  The PDF for $v_r$ at the base of the convection zone for case B
(Figure \ref{fig:hres_hist}$b$, red line) is also close to that
of the ASH simulation (black line).  Interestingly, the high power wings
present in case A in the upper portion of the domain are still present in
the PDF of case B (not shown).  These appear to be related instead to an
overdriving of the FSAM systems relative to ASH that arises from removal
of the polar regions, a subtle effect that we discuss shortly.

\subsection{Mean Flows and Thermodynamics of the System}
\begin{figure}
\centering
\includegraphics[width=3.0truein]{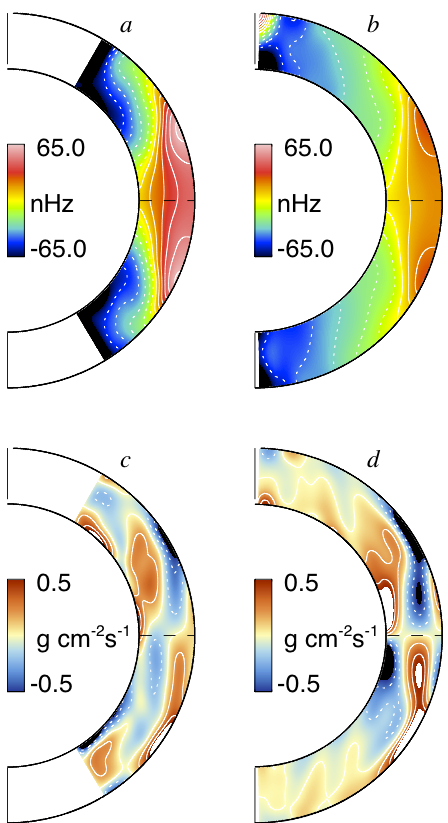}
\caption{Mean flows averaged in longitude and over 10 rotation periods.
\textit{(a)}  Differential rotation as realized with FSAM case A
and \textit{(b)}
ASH.  The differential rotation established in FSAM is somewhat stronger
than that realized in ASH.  \textit{(c)} Latitudinal mass flux achieved in
FSAM and \textit{(b)} ASH.  Blue (red) tones indicated poleward flow in the
northern (southern) hemisphere, while red (blue) tones indicate equatorward
flow.  Individual hemispheres tend to be dominated by a large circulation
cell for each simulation.
\label{fig:azavgs}}
\end{figure}

Convective flows realized in FSAM case A and ASH possess mean components
that are similar in nature to one another.  Figure \ref{fig:azavgs} depicts
the mean differential rotation and meridional circulations from each
simulation.  Case A exhibits a prograde differential rotation at the equator,
similar to that of ASH.  Meridional circulations in each case are
predominantly poleward in the upper convection zone and equatorward at the
base of the convection zone.  Closer inspection reveals that both simulations
tend to develop small counter cells of circulation in the near-surface
equatorial regions.  The differential rotation of case A, however, is
noticiably stronger than that realized in ASH.  This enhanced differential
rotation realized in FSAM is consistent with its convection being driven more
strongly than that of ASH, as suggested by the velocity PDFs.  As convection
becomes more vigorous, the resulting banana cells become more efficient at
establishing a prograde equator in systems such as these where the rotational
influence is strong.

\begin{figure}
\centering
\includegraphics[width=3.0truein]{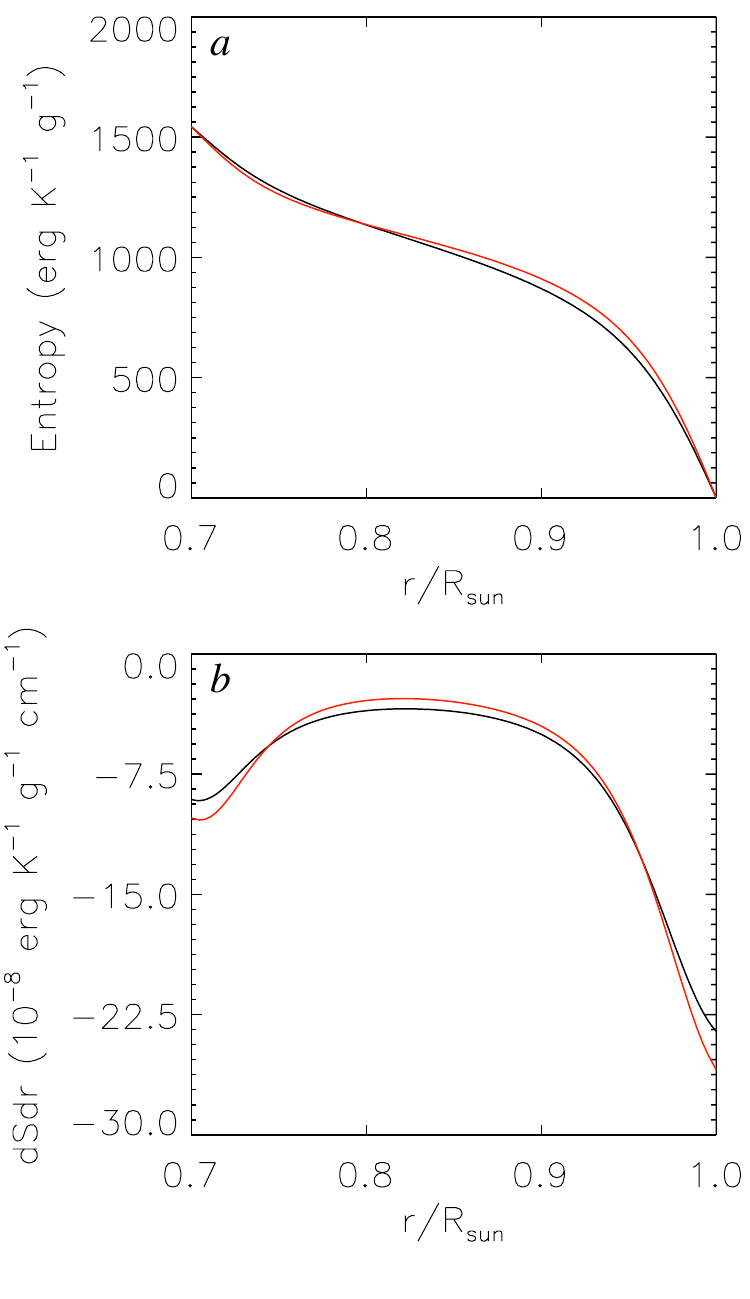}
\caption{Mean entropy profiles \textit{(a)} and entropy gradients \textit{(b)}
established in each case.  ASH results are plotted in black,
and FSAM case A in red.
Convection in the FSAM simulations tends to build a somewhat more adiabatic
interior and steeper entropy gradients near the boundaries than ASH.
\label{fig:sprofiles}}
\end{figure}

The disparity in convective driving becomes evident when looking at the
thermodynamic properties of these systems.  Figure \ref{fig:sprofiles}$(a)$
depicts the time-averaged, spherically symmetric entropy perturbations
attained by each simulation.  The profiles are similar, but convection in
case A (red) tends to build steeper gradients in the boundary layers than
that of ASH (black).  This is more readily apparent when looking at the
entropy gradient (Figure \ref{fig:sprofiles}$(b)$) for the two simulations.
The flows established by FSAM tend to establish an entropy
gradient that is 10$\%$ stronger near both the top and bottom boundaries
than that established in ASH, while
maintaining a more nearly adiabatic interior throughout the bulk of the
convection zone.  An enhanced entropy gradient near the boundaries translates
directly into an increase in the thermal energy throughput of the system.

\subsection{The Role of The Polar Regions}
How might the difference in convective driving between the two systems arise?
For constant entropy boundary conditions, convection is allowed to set the
latidudinal profile of the heat flux at the boundaries.  In the presence of
rotation, convective transport is
preferentially more efficient at the high latitudes
\citep[e.g.][]{Elliott:etal:2000} where Coriolis constraints on radial
motions are weakest. With the polar
regions absent, convection in FSAM has less freedom to establish such
latitudinal asymetries, and therefore leads to stronger driving of the
convective motions at the lower latitudes, and subsequently to stronger
banana cell-like structures.

\begin{figure}
\centering
\includegraphics[width=3.0truein]{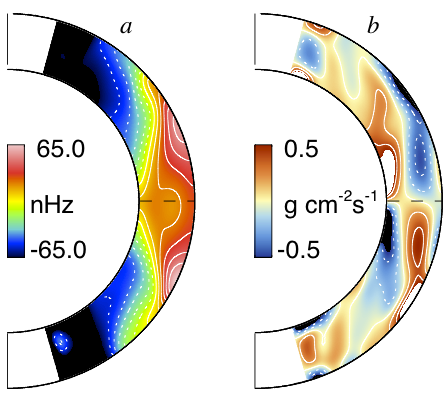}
\caption{Mean flows averaged in longitude and over 10 rotation periods for
FSAM case C.  \textit{(a)}  Differential rotation realized when a higher
latitude range is included shows reduced amplitude relative to case A, yielding
better agreement with the ASH simulation.  \textit{(b)} Latitudinal component
of mass flux for case C.   Blue (red) tones indicated poleward flow in the
northern (southern) hemisphere, while red (blue) tones indicate equatorward
flow. Meridional flows in case C also show good agreement with ASH.
\label{fig:highl_azavg}}
\end{figure}

\begin{figure}[h]
\centering
\includegraphics[width=3.0truein]{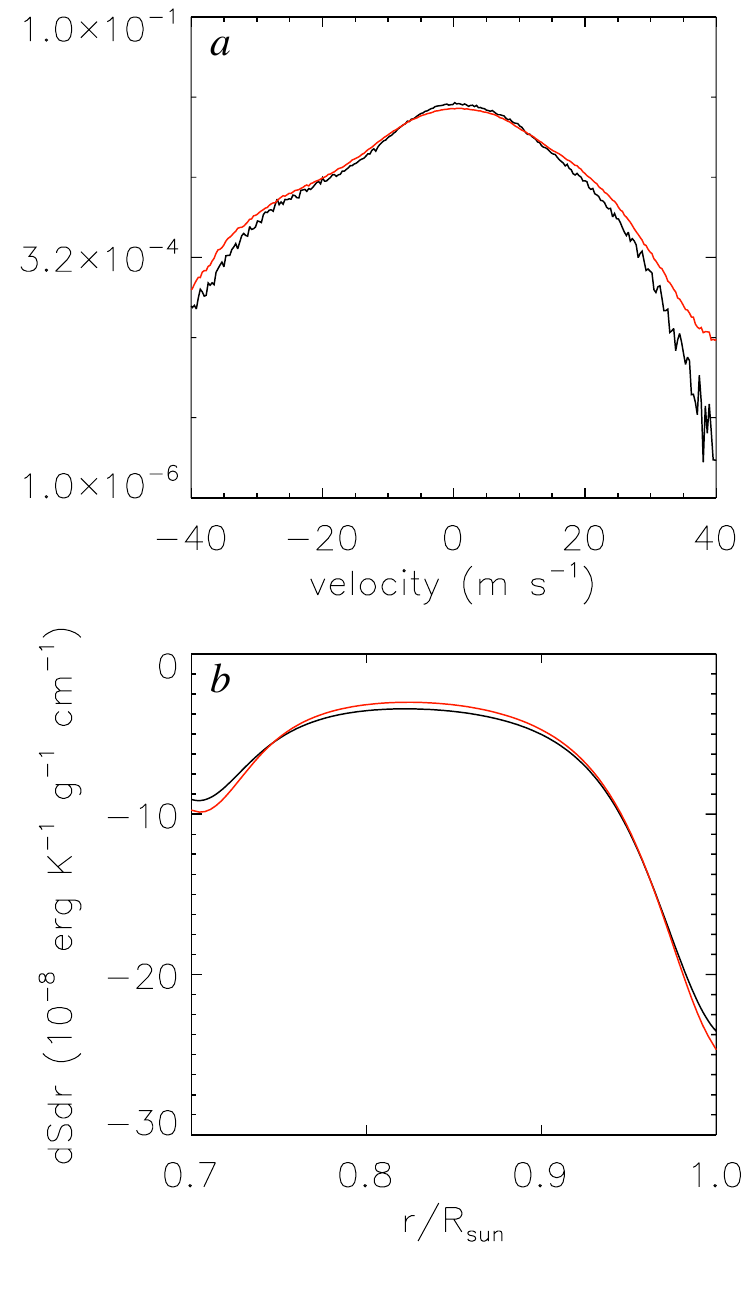}
\caption{\textit{(a)} PDFs of $v_r$ for FSAM case C (red) and the ASH case
(black) near the top of the simulation (0.99R).  The inclusion of higher
latitudes yields a PDF in case C that is in better agreement with the ASH
results than case A.  \textit{(b)} Mean entropy gradients established in
case C (red) and the ASH case (black).  The superadiabaticity of the boundary
layers present in case A is diminished as more latitudes are included in the
simulation.
\label{fig:highl_dsdr}}
\end{figure}

Extension of the latitudinal boundaries to $\pm$75$^{\circ}$, as we have
done with case C, allows us to examine this effect.  We find that in this
regime, FSAM results begin to converge toward those of ASH.
Figure \ref{fig:highl_azavg} depicts the mean flows for case C.  Meridional
circulations are very similar to those of the ASH simulation, and the
strength of the differential rotation, while still stronger than in ASH,
is diminished relative to case A.  The wings of the velocity PDF for
case C (Figure \ref{fig:highl_dsdr}$a$) have come down substantially from
their counterparts in case A, most noticiably so in the downflows.
Moreover, the steep entropy gradients that developed near the boundaries
of case A have diminished in case C relative to case A
(Figure \ref{fig:highl_dsdr}$b$.)

These tests suggest that FSAM, when properly resolved, can produce convective
flows in accord with those produced by the more widely used ASH code.  We
find reasonable convergence between the full- and partial-sphere simulations
as the latitudinal extent of the simulation is increased.  On the other hand,
these tests suggests that we may be well-cautioned to carefully consider the
luminosity we adopt for our simulations.  Otherwise we may inadvertently
overdrive the convection.  As the level of turbulence is increased, however,
convection tends to become more homogeneous in latitude
\citep[e.g.][]{Gastine:etal:2012, Featherstone:etal:2013}
 and we expect the role of the absent
poles to be diminished in the more turbulent regimes.

\section{Buoyant Rise of Active Region Flux Tubes in a Solar Like Convective Envelope \label{sec:tubes}}

\subsection{A simulation of rotating solar convection \label{sec:convsim}}
We now proceed to carry out a hydrodynamic simulation to obtain a
statistically steady
solution of a solar like, rotating convective flow field in a spherical shell
domain with $r \in [r_i, r_o]$, spanning from $r_i=0.722 R_{\odot}$ at
the base of the convection zone (CZ) to $r_o = 0.971 R_{\odot}$
at about 20 Mm below the photosphere, 
$\theta \in [ \pi/2 - \Delta \theta, \pi/2 + \Delta \theta]$ with
$\Delta \theta = \pi / 3$, and $\phi \in [0, 2 \pi]$.
The domain is resolved by a grid with 96 grid points in $r$,
512 grid points in $\theta$, and 768  grid points in $\phi$.
The grid is uniform in $r$, $\theta$, and $\phi$ respectively.
J. Christensen-Dalsgaard's solar model \citep{JCD:etal:1996},
commonly known as Model S,
is used for the reference profiles of $T_0$, $\rho_0$, $p_0$, $g_0$ in the
simulation domain. We assumed that $s_0 =0$ for the reference state, i.e. is
isentropic. We also omit the heating term due to radiative diffusion
$\nabla \cdot {\bf F}_{\rm rad}$ in the CZ in 
equation (\ref{eq:entropy1solv}), but
instead, drive convection by imposing at the lower boundary a fixed
entropy gradient $\partial s_1/\partial r$ such that the solar luminosity
$L_s$ is forced through the lower boundary as a diffusive heat flux:
\begin{equation}
\left ( K \rho_0 T_0 \frac{\partial s_1 }{\partial r} \right )_{r_i} = 
\frac{L_s}{4 \pi {r_i}^2}
\label{eq:lowbcs1}
\end{equation}
where $K = 2 \times 10^{13} \, {\rm cm}^2 \, {\rm s}^{-1}$
in our simulation domain.
We also impose a latitudinal variation of entropy at the lower boundary:
\begin{equation}
\left ( \frac{\partial s_1 }{\partial \theta} \right )_{r_i} =
\frac{d s_i (\theta) }{d \theta}
\label{eq:lowbcs1_lat}
\end{equation}
where
\begin{equation}
s_i (\theta) = - \Delta s_i \cos \left ( \frac{\theta - \pi / 2}
{\Delta \theta} \pi \right ) ,
\end{equation}
to represent the tachocline induced entropy variation that can break the
Taylor-Proudman constraint in the convective envelope.
In the above we set $\Delta s_i = 215.7 \: {\rm erg} \:
{\rm g}^{-1} \: {\rm K}^{-1}$.
For the initial condition, we let the initial $s_1$ be:
\begin{equation}
s_1 |_{t=0} = <\!\! s_1 \!\! >_{t=0} + s_i (\theta) - < \!\! s_i (\theta) \!\! >
\end{equation}
where $< \, >$ denotes the horizontal average at a constant $r$, and
$< \!\! s_1 \!\! >_{t=0}$ is given by:
\begin{equation}
K \rho_0 T_0 \frac{d < \!\! s_1 \!\! >_{t=0} }{d r} = 
\frac{L_s}{4 \pi r^2} ,
\label{eq:inits1}
\end{equation}
such that initially the constant solar luminosity is being carried through
the solar convection zone by thermal diffusion.
This results in an unstable initial stratification, and with a small initial
velocity perturbation, convection ensues in the domain.
For the upper boundary
$s_1$ is held fixed to its initial value, while at the lower boundary
the fixed gradient of $\partial s_1 / \partial r$ given by
equation (\ref{eq:lowbcs1}) maintains a conductive heat flux
corresponding to the solar luminosity through the lower boundary.
The latitudinal gradient of $s_1$ given by equation (\ref{eq:lowbcs1_lat}) is
also imposed at the lower boundary, but the horizontally averaged value of
$s_1$ is allowed to change with time. At the two
$\theta$ boundaries, $s_1$ is assumed symmetric.  The velocity field is
non-penetrating and stress free at the top, bottom and the two
$\theta$-boundaries.  The top and bottom boundary condition for
$p_1$ is
\begin{equation}
\rho_0 \frac{\partial}{\partial r} \left ( \frac {p_1}{\rho_0} \right )
= {\cal F}_r
\end{equation}
at $r= r_i$ and $r_o$, and ${\cal F}_r$ is the $r$-component of ${\cal F}$
given in equation (\ref{eq:rhs_pressuresolv}). At the two $\theta$-boundaries
\begin{equation}
\frac{\rho_0}{r} \frac{\partial}{\partial \theta}
\left ( \frac {p_1}{\rho_0} \right ) = {\cal F}_{\theta},
\end{equation}
and ${\cal F}_{\theta}$ is the $\theta$-component of ${\cal F}$
given in equation (\ref{eq:rhs_pressuresolv}).
All quantities are naturally periodic at the $\phi$ boundaries.
The kinematic viscosity $\nu = 10^{12} \, {\rm cm}^2 \, {\rm s}^{-1}$ in
the simulation domain. This gives a Prandlt number of $Pr = 0.05$ for our
simulation. The reference frame rotation rate $\Omega$ in
equation (\ref{eq:momentumsolv1}) is set to
$2.7 \times 10^{-6} \, {\rm rad} \, s^{-1}$, and with
respect to this frame, the initial velocity is essentially zero with
a very small initial perturbation.

\begin{figure}
\epsscale{1.}
\plotone{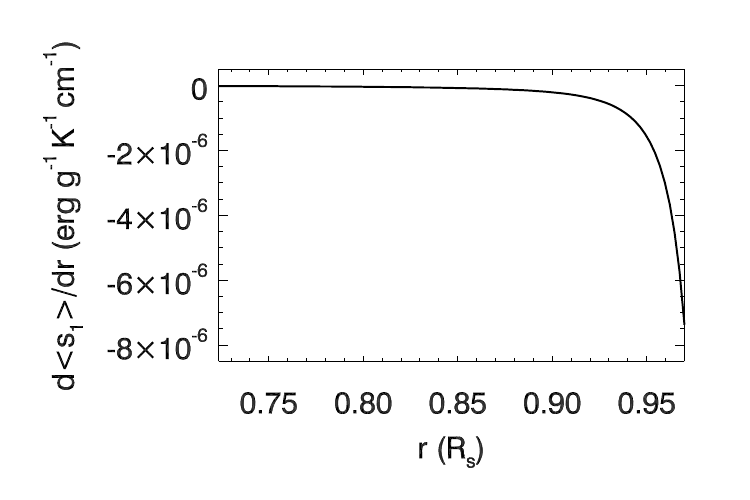}
\caption{The final steady state entropy gradient reached by the rotating
solar convective envelope in the convection simulation described in
Section \ref{sec:convsim}}
\label{fig:dsdrmean}
\end{figure}

\begin{figure}
\epsscale{1.}
\plotone{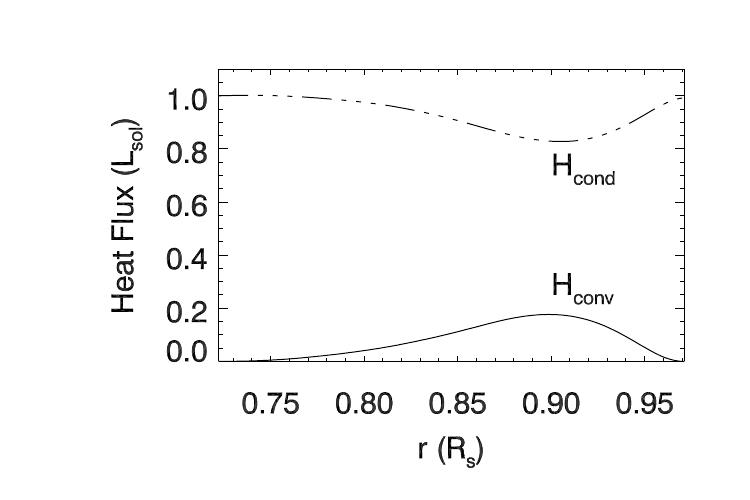}
\caption{Total heat flux due to
convection ($H_{\rm conv}$) and conduction ($H_{\rm cond}$) through the
convective envelope when the solution has reached a statistical steady
state.}
\label{fig:convs_fluxes}
\end{figure}

With the above setup of the simulation, we let the convection in the
domain evolve to a statistical steady state, which is reached after about
$6000$ days.  The final steady state entropy gradient reached by the
rotating solar convective envelope is shown in Figure \ref{fig:dsdrmean}.
The horizontally averaged entropy gradient reaches a value
of about $ - 7.5 \times 10^{-6} \, {\rm erg} \, {\rm K}^{-1} \, {\rm cm}^{-1}$
near the top boundary at about $0.97 R_s$, which is of a similar order of
magnitude as the entropy gradient
($\sim 10^{-5} \, {\rm erg} \, {\rm K}^{-1} \,
{\rm cm}^{-1}$) at this depth obtained by Model S \citep{JCD:etal:1996}.
Figure \ref{fig:convs_fluxes} shows the steady state, horizontally integrated
total heat flux due to convection:
\begin{equation}
H_{\rm conv} = \frac{4 \pi r^2}{A(r)}
\int_{0}^{2 \pi}
\int_{\pi/2 - \Delta \theta}^{\pi/2 + \Delta \theta}
\rho_0 T_0 v_r s_1 \, r^2 \, d \theta d \phi
\label{eq:Hconv}
\end{equation}
and conduction
\begin{equation}
H_{\rm cond} = \frac{4 \pi r^2}{A(r)}
\int_{0}^{2 \pi}
\int_{\pi/2 - \Delta \theta}^{\pi/2 + \Delta \theta}
K \rho_0 T_0 \frac{\partial s_1}{\partial r} \, r^2 \, d \theta d \phi
\label{eq:Hcond}
\end{equation}
where $A(r)$ is the total area of the spherical surface at radius r
\begin{equation}
A(r)=
\int_{0}^{2 \pi}
\int_{\pi/2 - \Delta \theta}^{\pi/2 + \Delta \theta}
\, r^2 \, d \theta d \phi .
\label{eq:area}
\end{equation}
In the above, the total heat fluxes $H_{\rm conv}$ and $H_{\rm cond}$ have
been scaled up to the values for the area of a full sphere so that they can
be compared directly with the solar luminosity $L_s$.
Here the conductive heat flux represents the heat transport due to
turbulent diffusion by unresolved convection. 
It can be seen from Figure \ref{fig:convs_fluxes} that with the large value
of $K$ used, the solar luminosity is mostly carried through by thermal
conduction and the heat flux transported by the resolved convective flow is
only a small fraction ($\sim 20$\%) of the solar luminosity. 
In this way the convective flow speed for the resolved giant cell convection
is not too high, (even with a relatively low viscosity $\nu = 10^{12} \,
{\rm cm}^2 \, {\rm s}^{-1}$ used), so that the convective flow is sufficiently
rotationally constrained to allow the maintainance of a solar-like
differential rotation with faster equator than the polar regions
\citep[e.g][]{Featherstone:Miesch:2012}.
The relatively low viscocity is chosen so that the subsequent simulations 
of the buoyant rise of active region flux tubes are not in a too viscous
regime.

\begin{figure}
\centering
\includegraphics[width=0.4\linewidth]{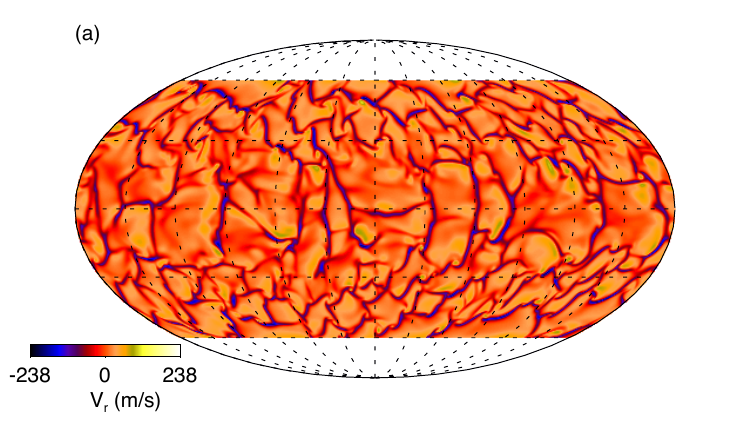}
\includegraphics[width=0.18\linewidth]{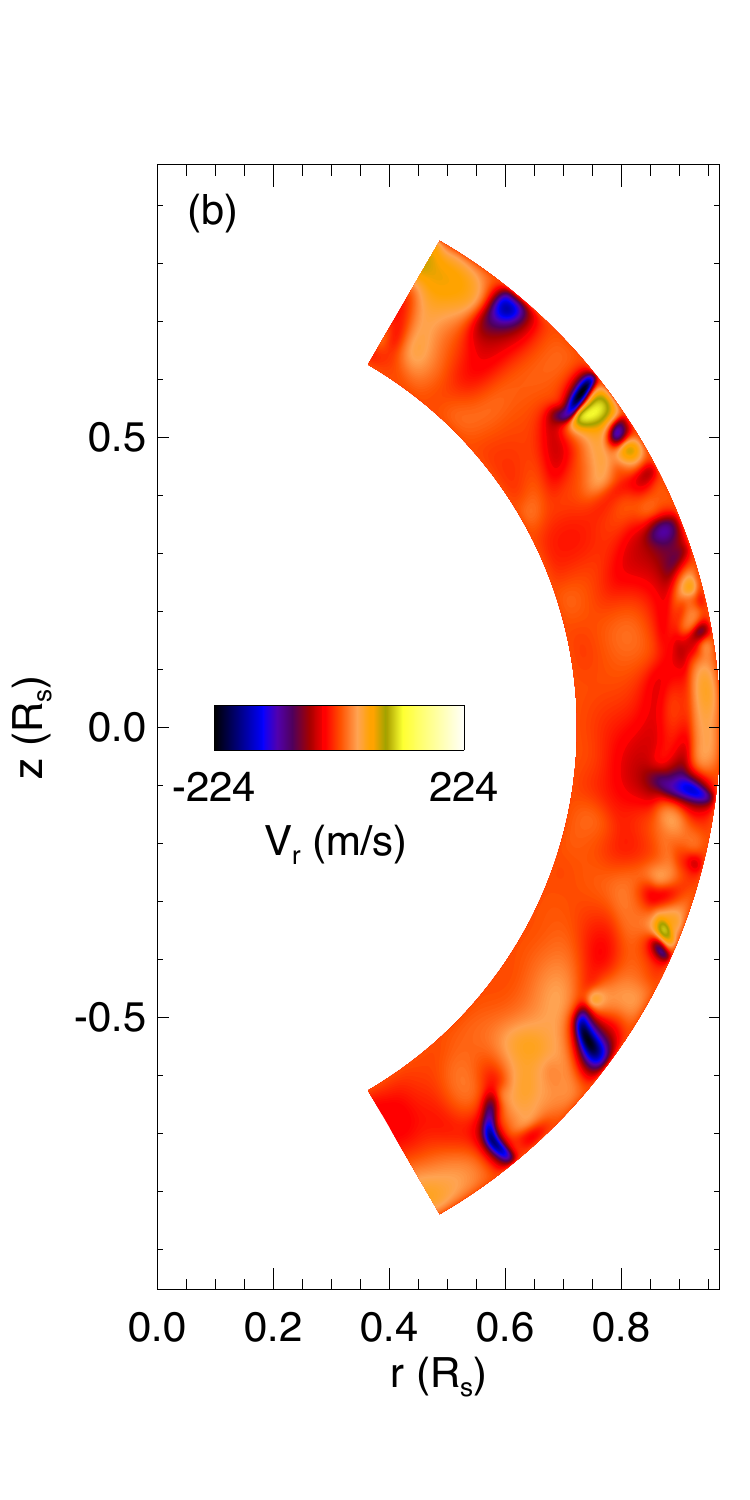}
\includegraphics[width=0.18\linewidth]{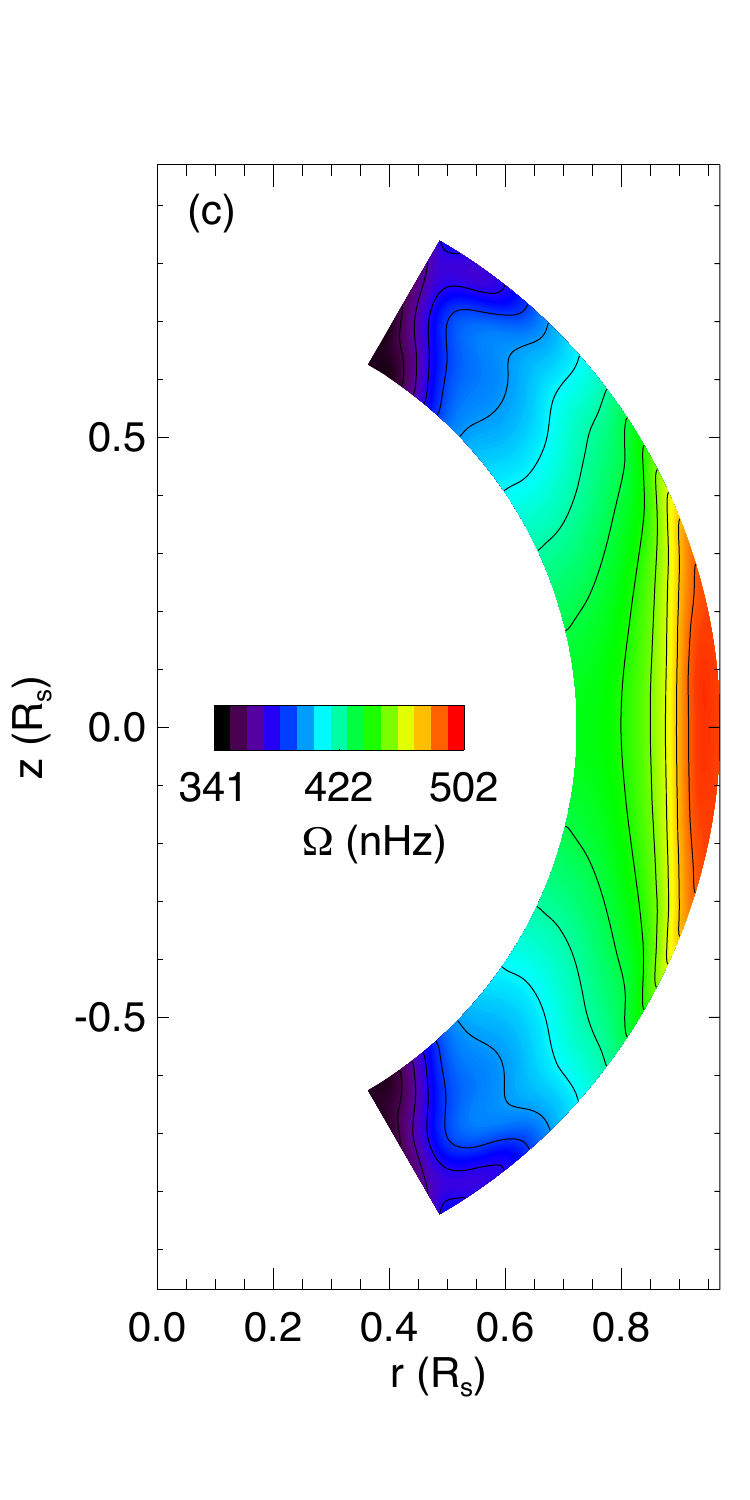}
\includegraphics[width=0.18\linewidth]{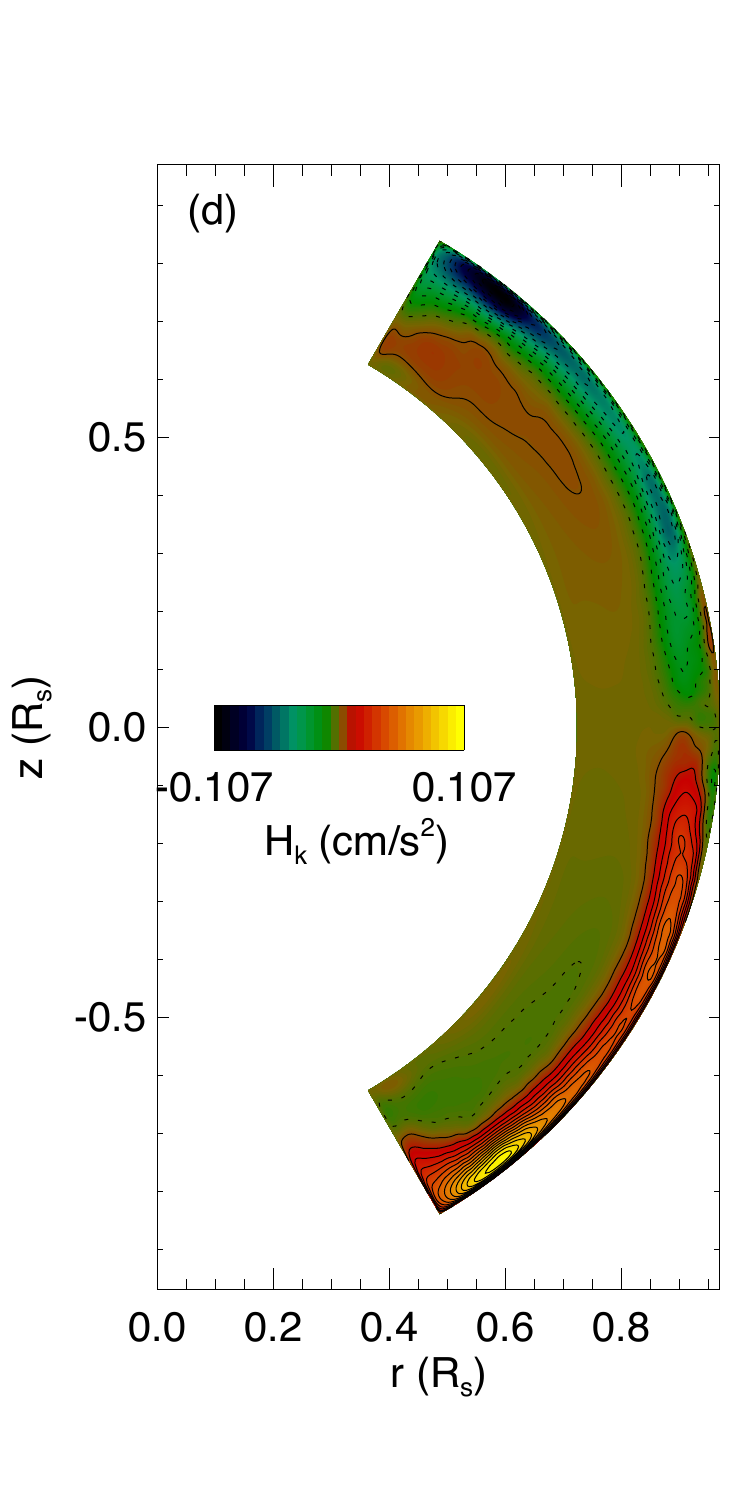}
\caption{(a) A snapshot of the radial velocity of the rotating solar
convection at a depth of about 30 Mm below the photosphere, shown on the
full sphere in Mollweide projection. (b) A meridional slice of the radial
velocity of the convective flow at the same time. (c and d) Time and
azimuthally averaged angular rate of rotation $\Omega$ and kinetic
helicity $H_k$ in the convective envelope after it has reached the statistical
steady state.}
\label{fig:mwdshell_omgmeri_hkmeri}
\end{figure}

Figure \ref{fig:mwdshell_omgmeri_hkmeri}a shows a snapshot 
of the radial velocity field of the rotating solar convection at a depth 
of about 30 Mm below the photosphere displayed on the full sphere in
Mollweide projection. It shows giant-cell convection patterns with
broad upflows in the network of narrow downflow lanes,
and with columnar, rotationally aligned cells (banana cells) at low latitudes.
The time and azimuthally averaged mean zonal flow
(Figure \ref{fig:mwdshell_omgmeri_hkmeri}c) shows a solar-like differential
rotation profile with faster rotation at the equator and slower rotation
towards the poles, and more conical shaped contours of constant angular
speed of rotation at mid-latitude range.
Figure \ref{fig:mwdshell_omgmeri_hkmeri}d shows the time
and azimuthally averaged kinetic helicity $H_k = {\bf v} \cdot (\nabla \times
{\bf v})$ of the flow.  It shows predominantly negative (positive) kinetic
helicity in the upper 1/3 to 1/2 of the CZ in the northern
(southern) hemisphere and weakly positive (negative) kinetic helicity in the
deeper depths of the CZ.
The depth of the upper layer with predominantly negative (positive) $H_k$ in
the northern (southern) hemisphere is relatively shallow because, as
can be seen in Figure \ref{fig:mwdshell_omgmeri_hkmeri}b, the
concerntrated downflow plumes do not penetrate very deep.
They generally
reach less than half of the total depth of the CZ before
starting to diverge and leading to a reversal of the kinetic helicity.

\subsection{Simulations of Rising Flux Tubes \label{sec:risetubesim}}

\subsubsection{Simulation Setup \label{sec:risetubesetup}}
Into the statistically-steady, rotating convective flow with a
self-consistently maintained solar like differential rotation, we insert
buoyant toroidal flux tubes near the bottom of the CZ to study how they
rise through the CZ. The initial flux tube we insert into the convecting
domain is given by the following:
\begin{equation}
{\bf B} = \nabla \times \left ( {A(r,\theta)
\over r \sin \theta } \hat{\bf \phi} \right )
+ B_{\phi} (r, \theta) \hat{\bf \phi},
\end{equation}
where
\begin{equation}
A(r,\theta) = {1 \over 2} q a^3 B_t
\exp \left( - {\varpi^2(r,\theta) \over a^2} \right),
\end{equation}
\begin{equation}
B_{\phi} (r, \theta) = {a B_t \over r \sin \theta}
\exp \left( - {\varpi^2(r,\theta) \over a^2} \right).
\end{equation}
\begin{equation}
\varpi = (r^2 + r_0^2 -2r r_0 \cos (\theta-\theta_0))^{1/2}.
\end{equation}
In the above, $q$ is the rate of twist (angle of field line rotation about
the axis per unit length of the tube), $a$ denotes the e-folding radius of
the tube, $r_0$ and $\theta_0$ are respectively the initial $r$ and $\theta$
values of the tube axis. For all of the simulations of this paper, $a=6.7
\times 10^8 {\rm cm}$ which is about 0.12 times the pressure scale height
at the base of the solar convection zone, $r_0=5.2 \times 10^{10} {\rm cm}$ is
at approximately $0.757 R_{\odot}$, $\theta_0$ corresponds to $15^{\circ}$
latitude, and the initial
field strength at the axis of the toroidal flux tube is $B_t a/( r_0
\sin \theta_0 ) = 10^5 {\rm G}$. Thus the total flux in the initial toroidal
flux tube is $1.4 \times 10^{23}$ Mx, which is about a factor of 10 greater
than the typical flux in a solar active region.  Due to the limited numerical
resolution of our global scale simulations of the convective envelope, we
can only consider tubes with a rather large cross-section in order for
it to be resolved by the numerical grid.  In our current simulations the
initial tube diameter is resolved by about 7 grid points.

\begin{figure}
\epsscale{0.8}
\plotone{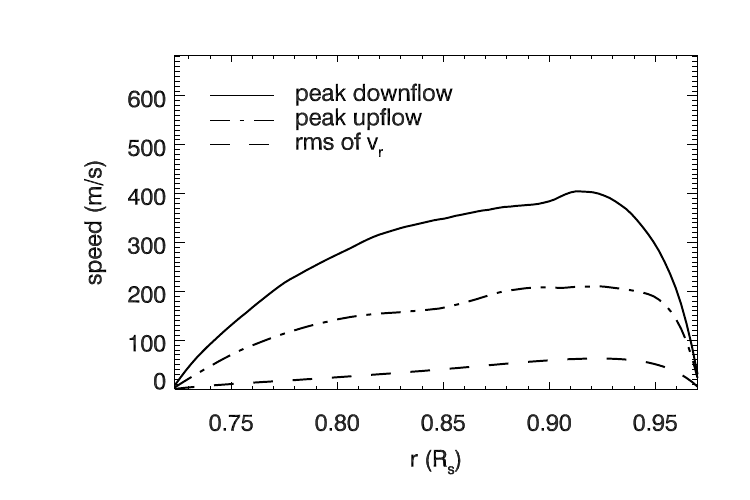}
\caption{Peak downflow speed, peak upflow speed, and the RMS vertical flow
speed of the convective flows in the solar convective envelope as a function
of depth, averaged over 30
evenly spaced temporal samples over a period of about 3 months.}
\label{fig:v1peakrms}
\end{figure}

We consider initially buoyant toroidal flux tubes, and specify the initial
buoyancy along the tube in the following two ways. 
In one way, an initial sinusoidal variation (with an azimuthal
wavelength of $\pi/2$ in $\phi$) of entropy:
\begin{equation}
\delta s_1 =c_p \left( 1- { 1 \over \gamma} \right )
{B_{\phi}^2 \over 8 \pi p_0}
\left [ {1 \over 2} \left ( 1 - {1 \over {\gamma - 1}} \right )
- {1 \over 2} \left ( 1 + {1 \over {\gamma - 1}} \right ) \cos (4 \phi)
\right ]
\label{eq:m4_buo}
\end{equation}
is being added to the original $s_1$ of the convective flow field at the
location of the toroidal tube.
Thus along each $\pi/2$ azimuthal segment of the toroidal tube, the tube
is varying from being (approximately) in thermal equilibrium with the
surrounding and thus buoyant, to being approximately in neutral buoyancy.
The peak buoyancy in the initial tube is approximately
$B_{\phi}^2 / 8 \pi H_{p0}$, corresponding to the
magnetic buoyancy associated with a flux tube in thermal equilibrium
with its surrounding.
Another initial buoyancy state we used is to specify a
uniform buoyancy along the tube, by adding
\begin{equation}
\delta s_1 =c_p \left( 1- { 1 \over \gamma} \right )
{B_{\phi}^2 \over 8 \pi p_0}
\label{eq:uniform_buo}
\end{equation}
to the original $s_1$ of the convective flow field at the location of
the toroidal tube.  In this way it is uniformly buoyant along the tube
with the magnetic buoyancy $B_{\phi}^2 / 8 \pi H_{p0}$.
We run two simulations of rising flux tubes in the convective flows with the
sinusoidal initial buoyancy (eq. [\ref{eq:m4_buo}]), one with a weak initial
(left-handed) twist rate of $q=-0.15 a^{-1}$, and the other with no initial
twist $q=0$. We name these runs ``SbWt'' (Sinusoidal-buoyancy-Weak-Twist) and
``SbZt'' (Sinusoidal-buoyancy-Zero-twist) respectively.
As a reference for these two simulations,
we run two corresponding simulations of the same initial buoyant tubes
in a quiescent rotating envelope with no convective flows,
but with the same reference stratification of $p_0$, $\rho_0$, and $T_0$.
These two runs are named ``SbWt-ref'' and ``SbZt-ref''.
Furthermore, we run a
simulation (named ``UbZt'') of the uniformly buoyant initial tube
(using eq. [\ref{eq:uniform_buo}]) with no initial twist rising in the
convective flow.  A summary of these runs is given in Table \ref{tubes_table}.
In this paper we only conduct these few sample runs to examine qualitatively
how a solar-like rotating convective flow may influence the rise of
relatively strong ($100$ kG) buoyant flux tubes.
The peak Alfv\'en speed $v_a$ in the initial flux tube is $764$ m/s.
Compared to the convective flow speeds
shown in Figure \ref{fig:v1peakrms}, the flux tube is significantly
super-equipartition with respect to the mean kinetic energy of the 
convective flows as reflected by the RMS velocity.  However, as discussed in
\citet{Fan:2003}, the hydrodynamic forces from the convective flows
would be able to counteract the magnetic buoyancy of the flux tube if the
speed of the convective flows is $\gsim (a/H_p)^{1/2} v_a$ which is
$\sim 265 $ m/s considering the initial radius $a$ of the buoyant toroidal
flux tubes in the present simulations.
Figure \ref{fig:v1peakrms} shows that the peak downflow speed
exceeds that value for most of the convection zone, indicating that the
downflow plumes can significantly impede the buoyant rise of the flux tube
even for the $100$ kG strong flux tubes considered here.

For the simulations of the rising flux tubes, we preserve the
kinematic viscosity $\nu = 10^{12} \, {\rm cm}^2 \, {\rm s}^{-1}$
used for the simulation of the rotating convective flow solution
in the entire simulation domain.  The thermal diffusivity used in the
original convection simulation is much greater ($2 \times 10^{13} \,
{\rm cm}^2 \, {\rm s}^{-1}$).  This large value is used in order to achieve
a solar like differential rotation profile (fast equator, slow poles) in
the rotating convection solution.  For the rising flux tube simulations, we
apply a magnetic field strength dependent quenching of $K$:
\begin{equation}
K=\frac{K_0}{1+(B/B_{\rm cr})^2}
\end{equation}
where $K_0 = 2 \times 10^{13} \, {\rm cm}^2 \, {\rm s}^{-1}$ is the
original value of the diffusivity used in the convection simulation and
$B_{\rm cr} = 100$ G represents a low threshhold field strength above which
quenching of thermal diffusion begins to take place.
Convection is expected to be suppressed by the strong magnetic field in the
flux tube, thus $K$, which represents unresolved eddy diffusion, should be
suppressed in the rising flux tubes.
For the magnetic field, we also do not include any explicit resistivity
$\eta$ in the simulation, so only numerical diffusion is present. This way
we minimize magnetic diffusion to preserve the frozen-in condition
of the buoyant flux tube as much as possible, given the numerical resolution.

\subsubsection{Results}
\begin{figure}
\epsscale{0.8}
\plotone{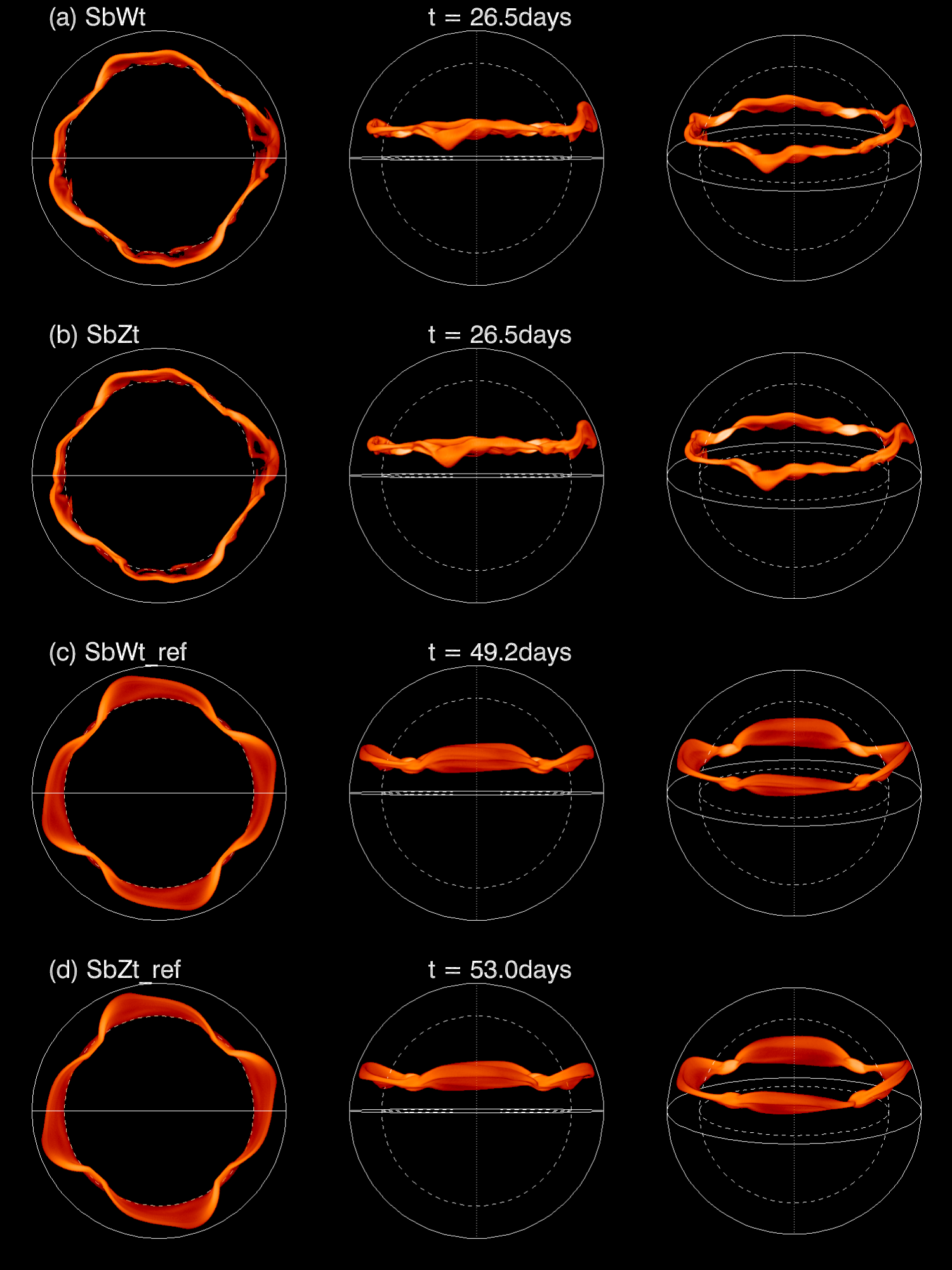}
\caption{3D volume rendering of the absolute magnetic field strength of
the rising flux tubes developed from simulations SbWt (a), SbZt (b),
SbWt-ref (c), and SbZt-ref(d) when an apex of the tube has reached the
top boundary. Animations of the evolution of the tube for each simulations
are available in the online version of the paper.}
\label{fig:3dtubes}
\end{figure}

Figures \ref{fig:3dtubes}a and \ref{fig:3dtubes}b show the rising flux tubes
that have developed from the SbWt and SbZt simulations respectively, when
an apex of the tube has reached the top boundary.  For comparison, the
resulting rising tubes from the corresponding reference simulations
SbWt-ref and SbZt-ref (without convection) are shown in
Figures \ref{fig:3dtubes}c and \ref{fig:3dtubes}d.
MPEG movies of the evolution of the tube for each of the simulations are
available in the online version of the paper.
In the absence of convection, four identical rising loops develop due to
the initial buoyancy prescription and rise to the top of the domain.
Convective flows are found to produce additional undulations on the rising
loops, pushing down certain portions while promoting the rise of other
portions.  With convection, the rise time for an apex of the tube to reach
the top is significantly reduced (for example, changed from about 49 days for
SbWt-ref to 26.5 days for SbWt).

\begin{figure}
\centering
\includegraphics[width=0.2\linewidth]{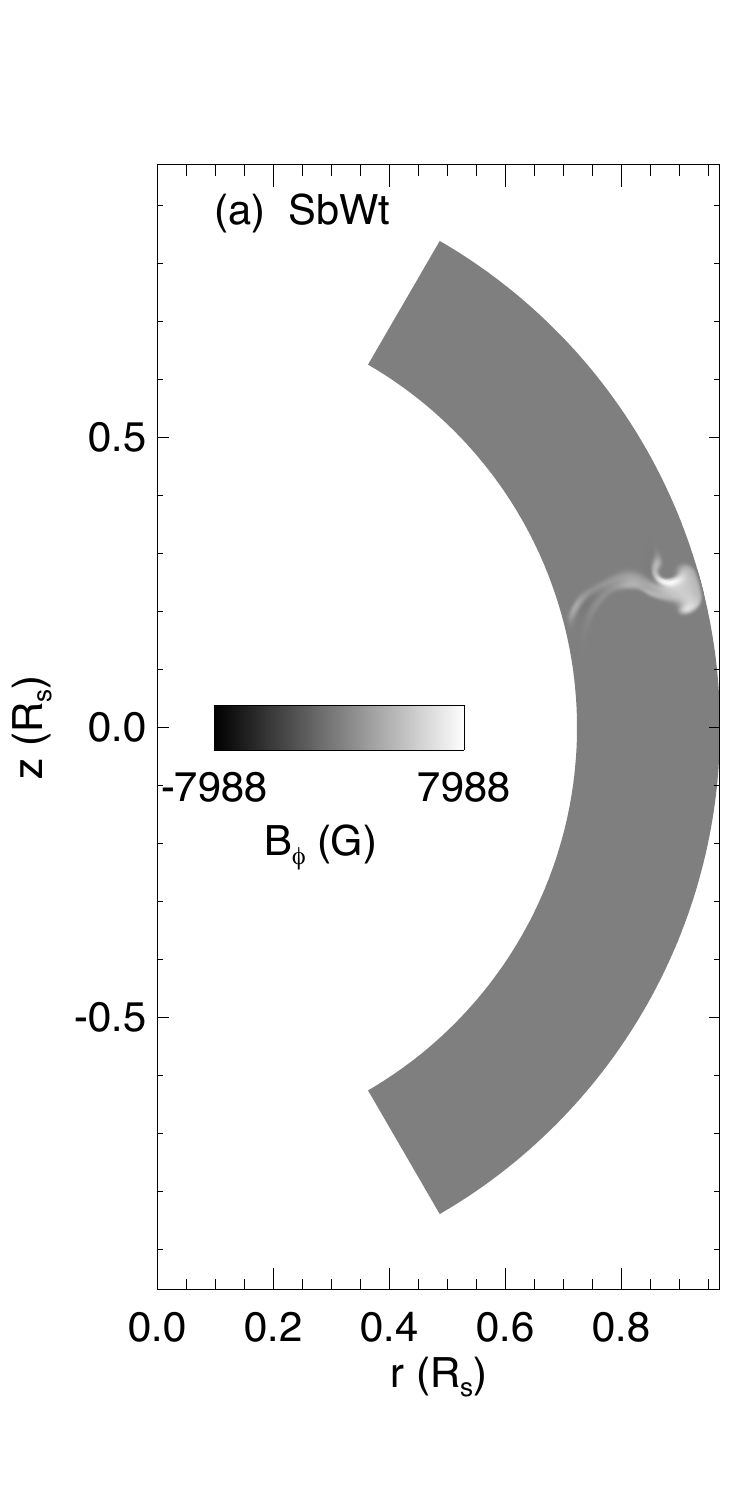}
\includegraphics[width=0.2\linewidth]{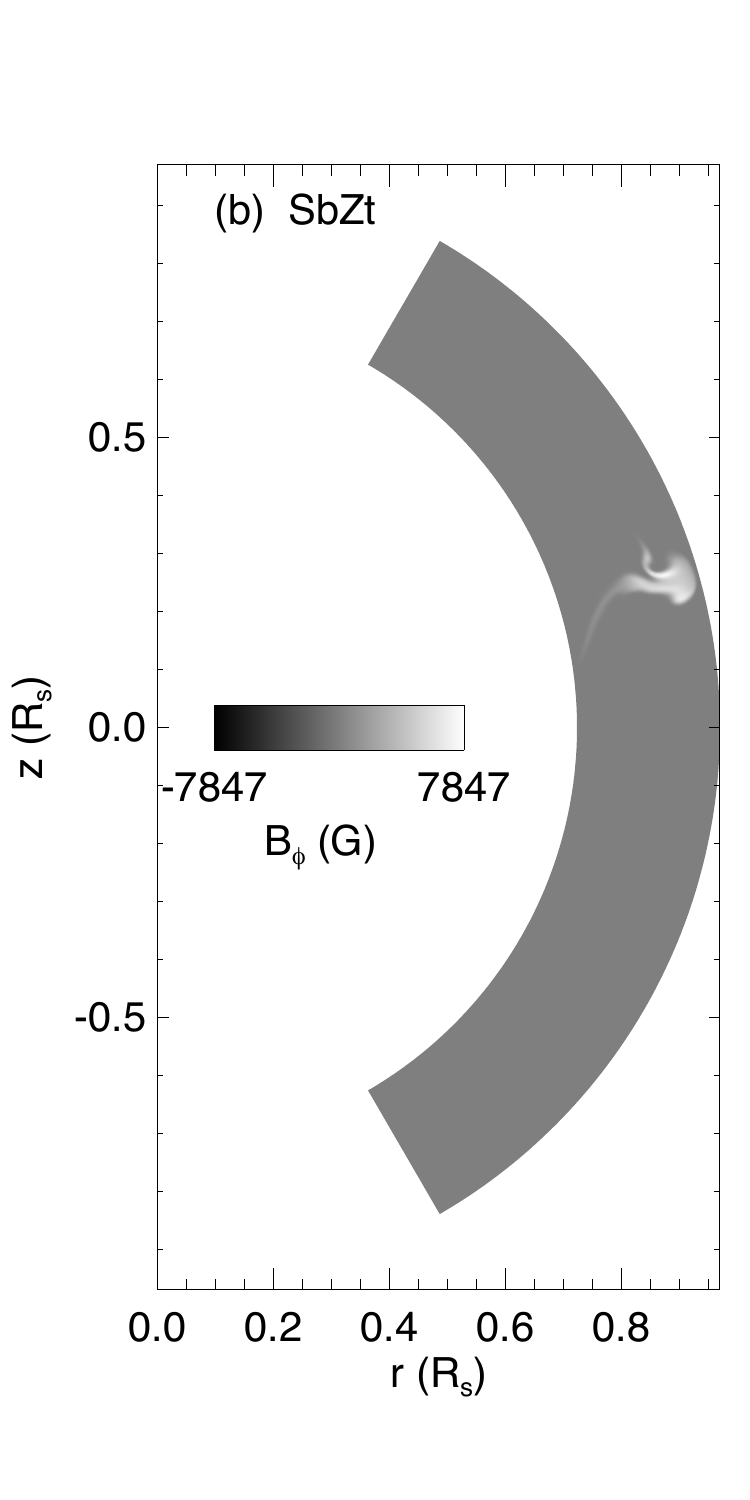}
\includegraphics[width=0.2\linewidth]{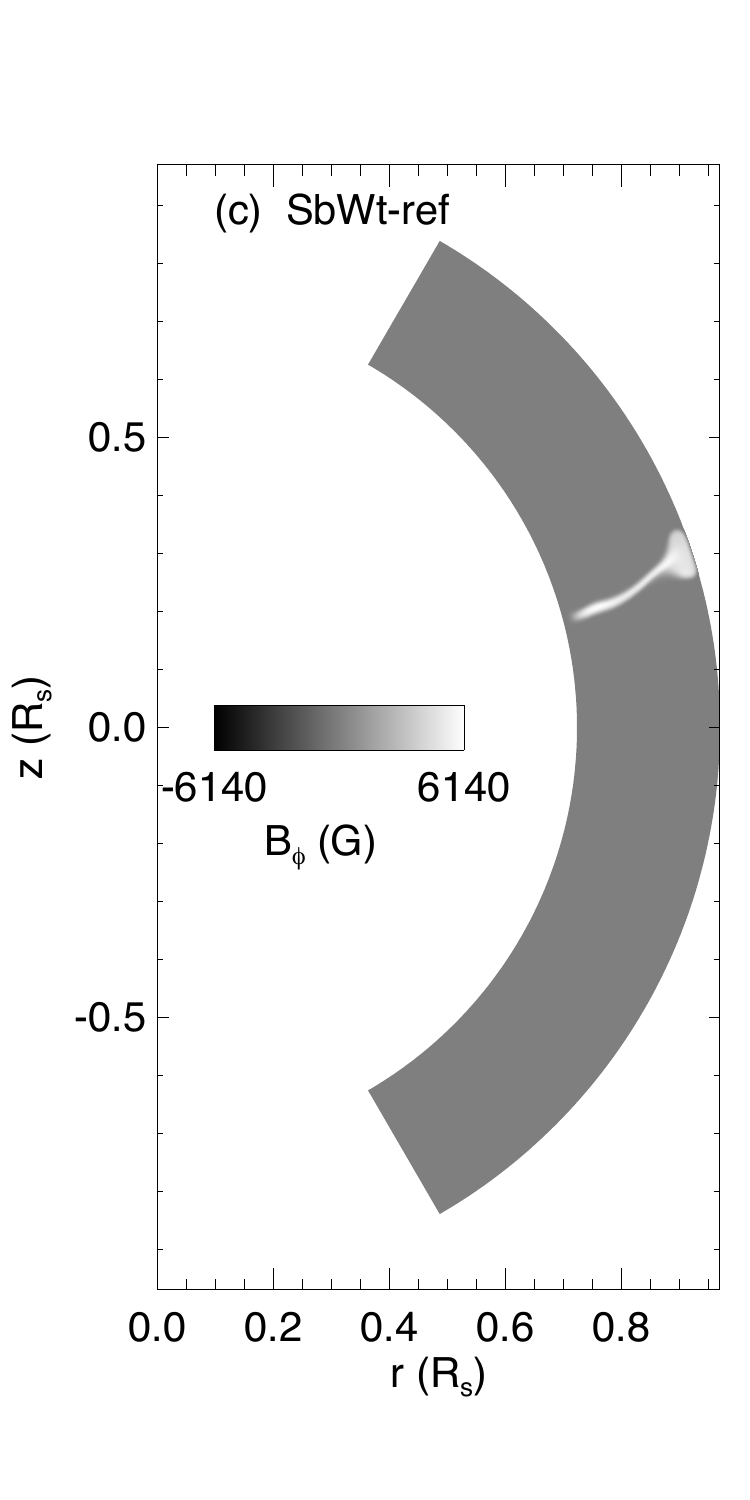}
\includegraphics[width=0.2\linewidth]{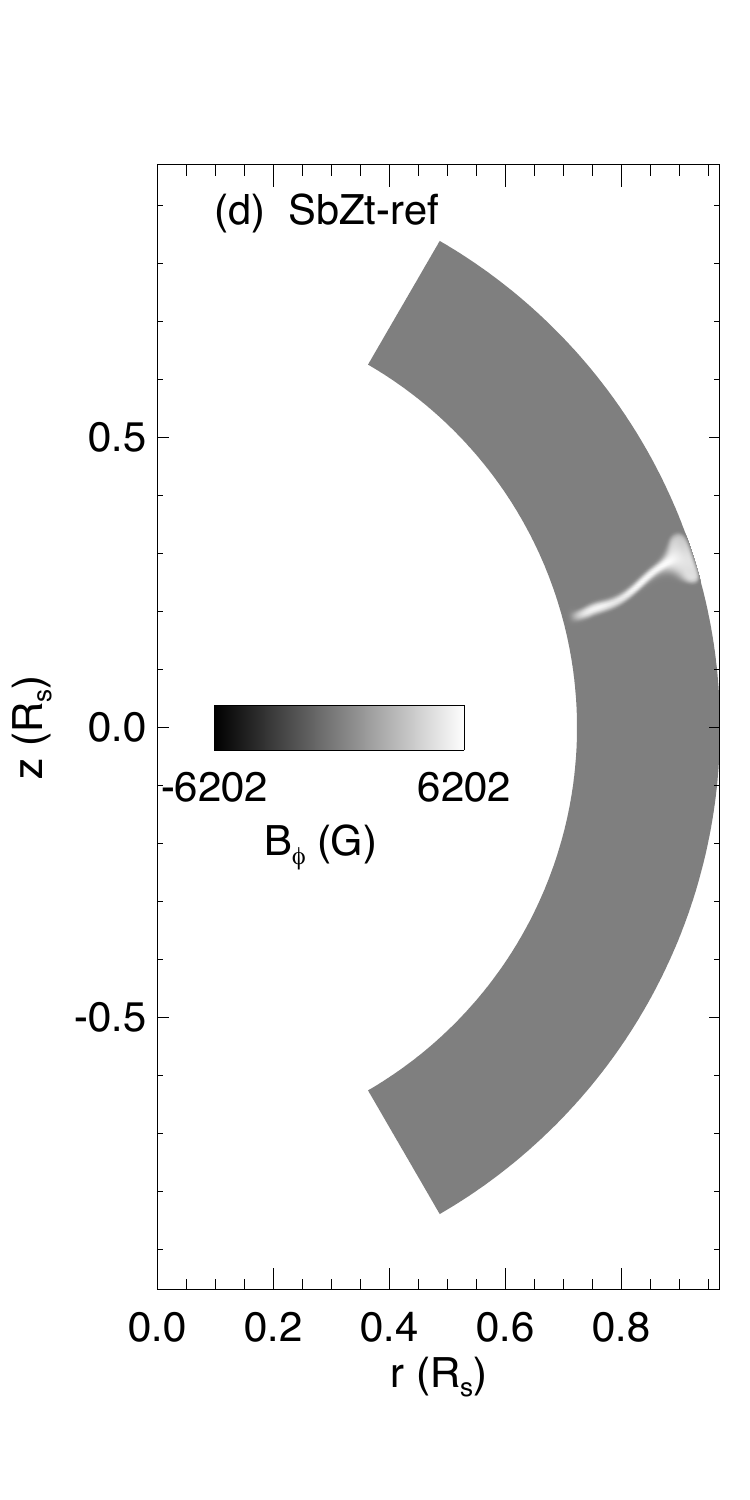}
\caption{$B_{\phi}$ in the meridional plane at the longitude of the
apex location for cases SbWt (a), SbZt (b) , SbWt-ref (c), and
SbZt-ref (d) at the same corresponding times shown in Figure \ref{fig:3dtubes}}
\label{fig:merislices}
\end{figure}

There is little difference in the morphology of the rising tubes (at least
as shown in the volume rendering of the absolute
magnetic field strength) between the weakly twisted and the untwisted cases,
both with and without convection.
One of the reasons for this is that the twist is rather weak, about a
half of the necessary twist rate
required for a cohesive rise of the flux of the
original flux tube as a whole, similar to the weakly twisted case
studied in \citet{Fan:2008} (see the LNT case shown in Figure 8 of that paper).
In other words, the magnetic energy density associated with the initial twist
component of the field (i.e. the $B_{\theta}$ and $B_r$ components in
the initial toroidal flux tube) is smaller than the kinetic energy density
associated with the relative velocity between the tube and the surrounding
plasma. As a result the initial twist does not have a great effect
on maintaining the cohesion of the rising tube
compared to the untwisted case.  There is continued flux loss during the
rise, forming a track of flux behind the rising apex,
as can be seen in the meridional cross-section of $B_{\phi}$
at the apex longitude for all the cases as shown in Figure \ref{fig:merislices}.
We also note that the current simulations of the rising flux tubes
are in a fairly laminar regime.  The Reynolds number for the rising flux tube
estimated based on the tube diameter $D \sim 10^9$ cm, typical
rise speed attained $V_{\rm rise} \sim 100 $ m/s, and the viscosity
$\nu = 10^{12} \, {\rm cm}^2/{\rm s}$ (kept the same as that used for
obtaining the
rotating convection solution), is $R_e = V_{\rm rise} D / \nu \sim 10$. 
Such a low Reynolds number reduces the production of small scale features
and fragmentation of the flux tube and thus generally improves the cohesion
for the rising flux, especially for the untwisted case.  This is also a
reason for the reduced difference in the magnetic field morphology between
the untwisted and weakly twisted cases.

\begin{figure}
\centering
\includegraphics[width=0.4\linewidth]{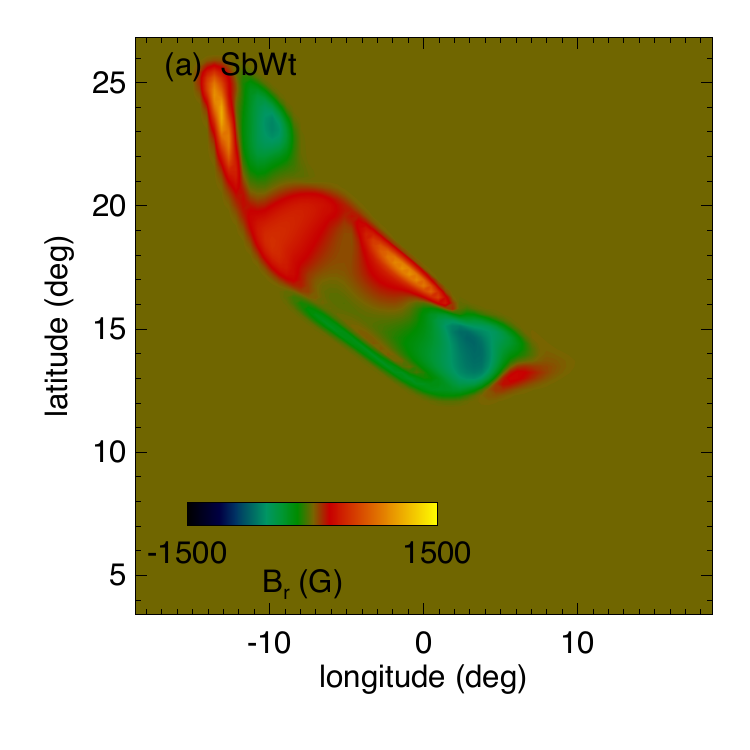}
\includegraphics[width=0.4\linewidth]{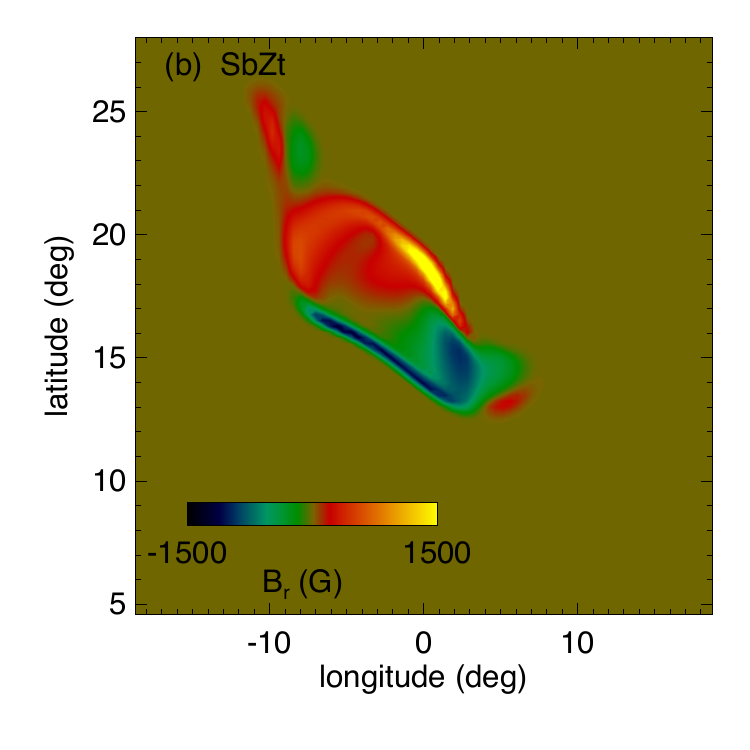}
\includegraphics[width=0.4\linewidth]{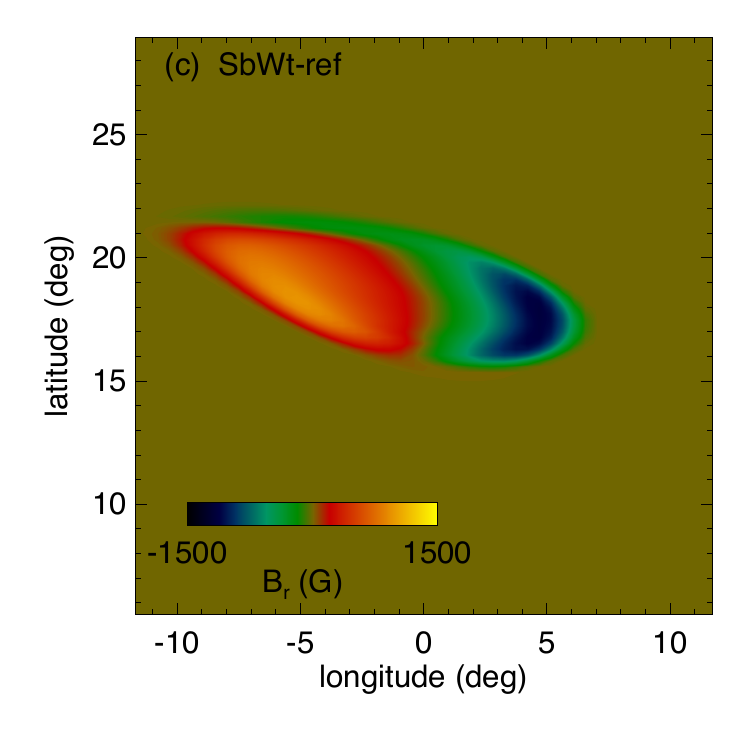}
\includegraphics[width=0.4\linewidth]{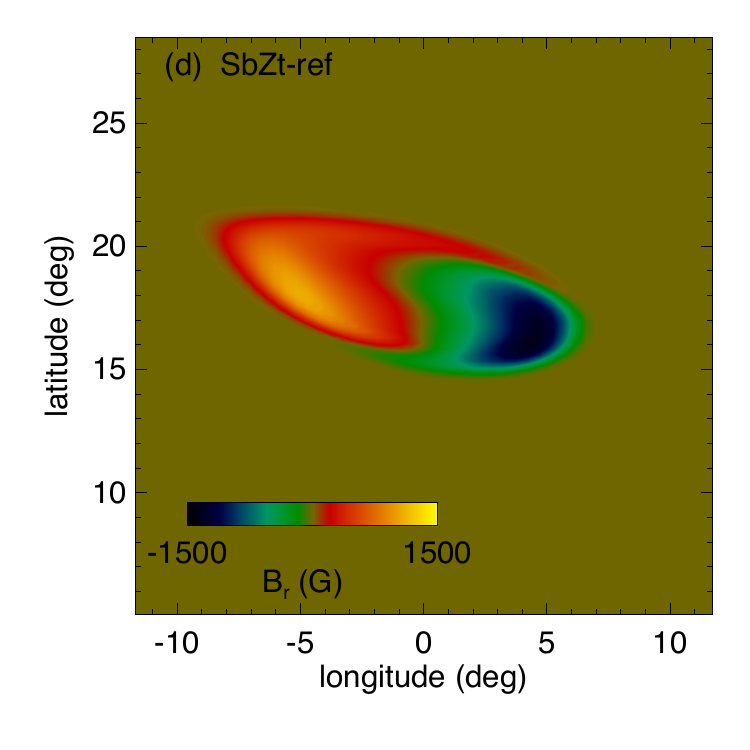}
\caption{The normal flux distribution produced by the top of the rising tube
approaching the upper boundary on a constant $r$ surface
at $r=0.957 R_s$, for the cases SbWt (a), SbZt (b), SbWt-ref (c), and
SbZt-ref (d).  Note in these plots, we have shifted the longitude of the
apex location of the rising flux to 0 degree longitude.}
\label{fig:emgfluxpattern}
\end{figure}

In all the cases, the apex rises nearly radially, with a small poleward drift.
Figure \ref{fig:emgfluxpattern} shows the normal flux distribution 
produced by the emerging apex portion near the top boundary on a constant
$r$ surface at $r=0.957 R_s$.
It can be seen that for all of the four cases the latitude of emergence is
centered at a location just slightly poleward
(by no more than about $3.5^{\circ}$) than the initial latitude of
$15^{\circ}$.  For the cases without convection (panels (c) and (d)), the
apex portion produces a simple bipolar structure with a tilt angle of
$7.2^{\circ}$ clockwise for the weakly twisted (SbWt-ref) case, and
$16^{\circ}$ clockwise for the untwisted (SbZt-ref) case.
These tilts are consistent with the mean tilt of solar active regions as
described by Joy's law.
With convection, the additional distortion and undulation
caused by the convective flows produce a more complex 
emergence pattern with multiple bipolar structures in the SbWt and SbZt
cases as shown in Figures \ref{fig:emgfluxpattern}a and
\ref{fig:emgfluxpattern}b.
However, the leading (negative) polarity flux is on average equatorward and
westward of the following (positive) polarity, consistent with the
direction of the active region mean tilt. The tilt angle
as determined by the flux weighted positions of the leading and following
polarity flux concentrations is $29.2^{\circ}$ clockwise for the
weakly twisted (SbWt) case and $53.2^{\circ}$ clockwise for the
untwisted (SbZt) case, which are of the right sign but are of a
significantly greater magnitude than the active region mean tilt.

\begin{figure}
\centering
\includegraphics[width=0.56\linewidth]{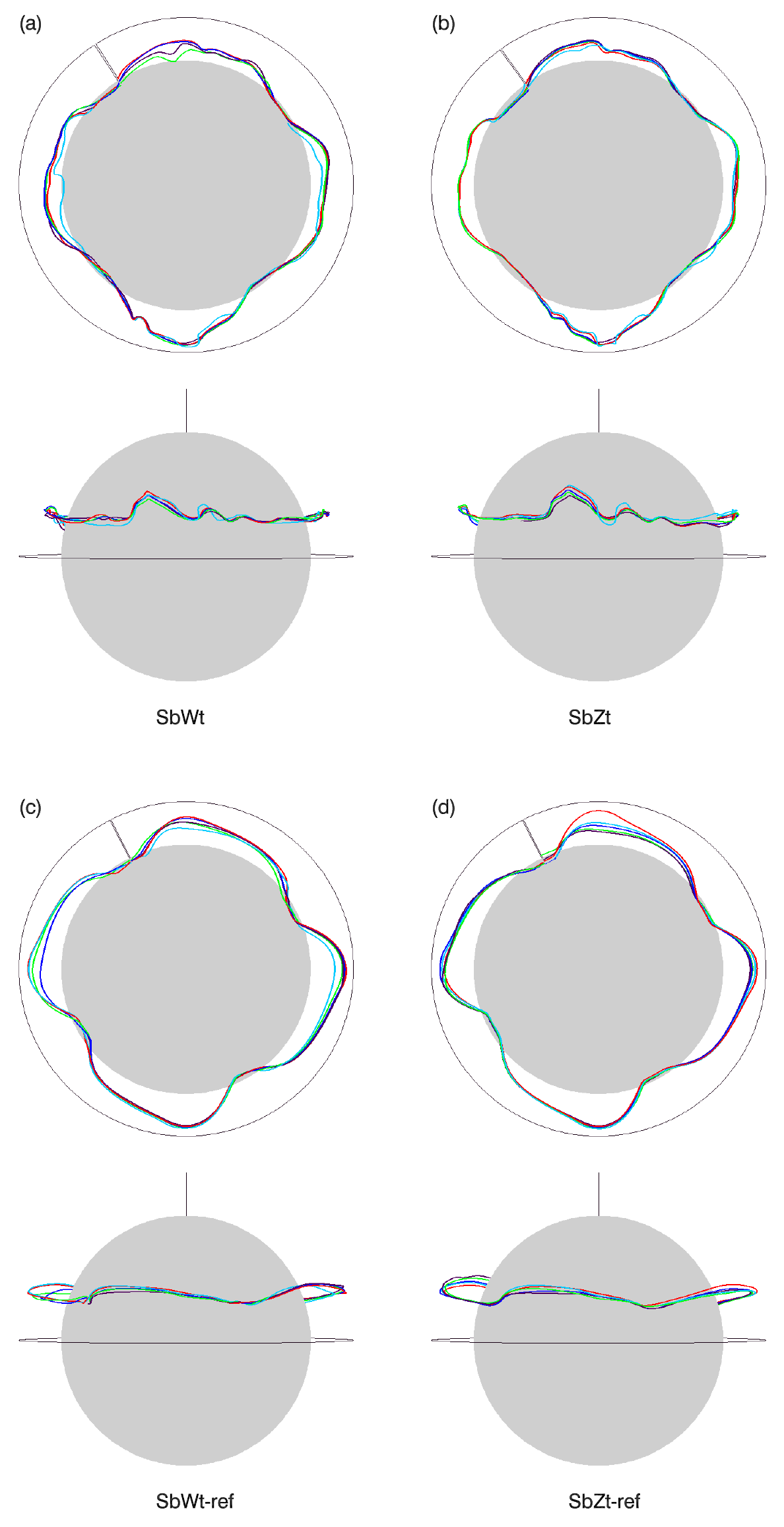}
\caption{Polar and equatorial views of selected field lines in the rising
flux tubes for the cases SbWt (a), SbZt (b), SbWt-ref (c), and SbZt-ref (d).
For each of the cases, the polar (equatorial) view is the upper (lower) panel}
\label{fig:3dfdls}
\end{figure}

Figure \ref{fig:3dfdls} shows 3D views of a few selected field lines traced
from the apex portion in the
rising flux tubes for the four cases: (a) SbWt, (b) SbZt, (c) SbWt-ref,
and (d) SbZt-ref, as viewed from the pole (upper panel in each case), and
from the equator (lower panel in each case).
For all the cases, the apex of the rising tube is at the 6 o'clock
location in the polar view and at the central meridian in the equatorial
view. It can be seen that the field lines at the apex are pointing
southeast-ward, i.e. consistent with the sense of tilts of solar active
regions. The tilt angles of the field orientation from the east-west
direction are significantly bigger in the convective cases (SbWt and SbZt
in Figures \ref{fig:3dfdls}a and \ref{fig:3dfdls}b)
compared to the non-convective reference cases (SbWt-ref and SbZt-ref
in Figures \ref{fig:3dfdls}c and \ref{fig:3dfdls}d). 
In these particular convective cases, the convective flows have driven
additional clock-wise rotation of the fields at the rising apex.
A statistical study with many more simulations of rising flux tubes,
sampling different times and locations of the convective flows (as
was done with the thin flux tube model in
\citet[e.g.][]{Weber:Fan:Miesch:2012})
are needed to determine whether the tilt angles at the apex of the emerging
flux obey Joy's law for solar active regions.
Our initial simulations here show that even with a relatively
strong initial magnetic field of 100 kG, a solar-like giant cell
convection can significantly reshape the buoyantly rising loops and
shorten the time for the apex to reach the top.

\begin{figure}
\centering
\epsscale{0.8}
\plotone{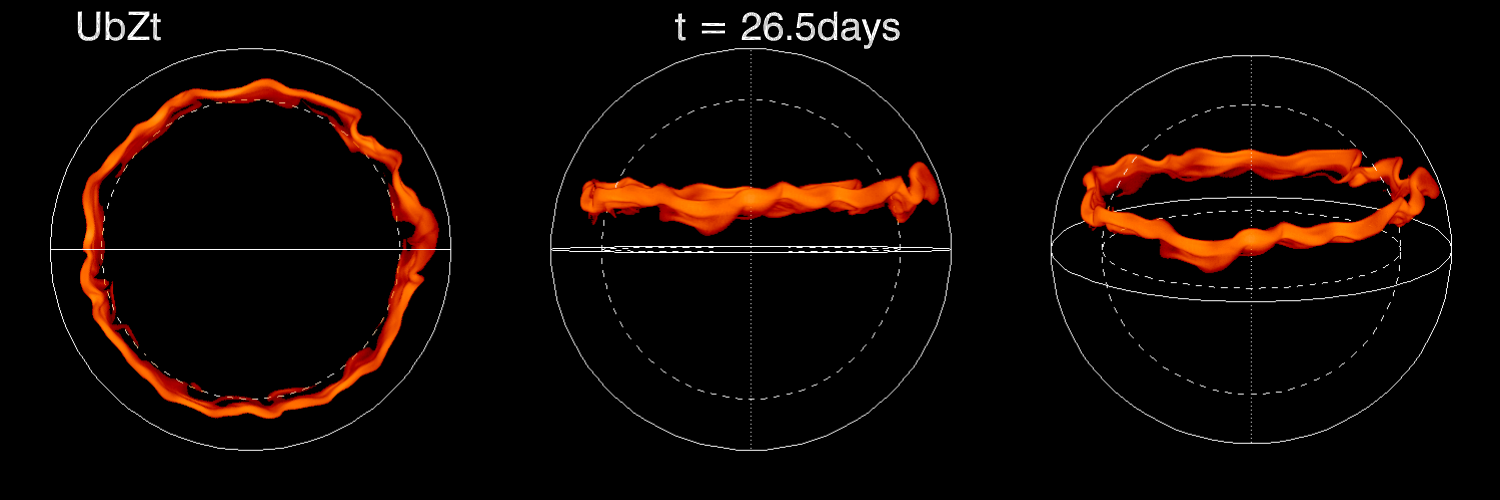}
\caption{3D volume rendering of the absolute magnetic field strength of the
rising flux tube developed from the UbZt simulation as the apex at the right
is reaching the top boundary. An MPEG animation of the evolution is available
in the online version.}
\label{fig:ubzt_3dtube}
\end{figure}

\begin{figure}
\centering
\includegraphics[width=0.28\linewidth]{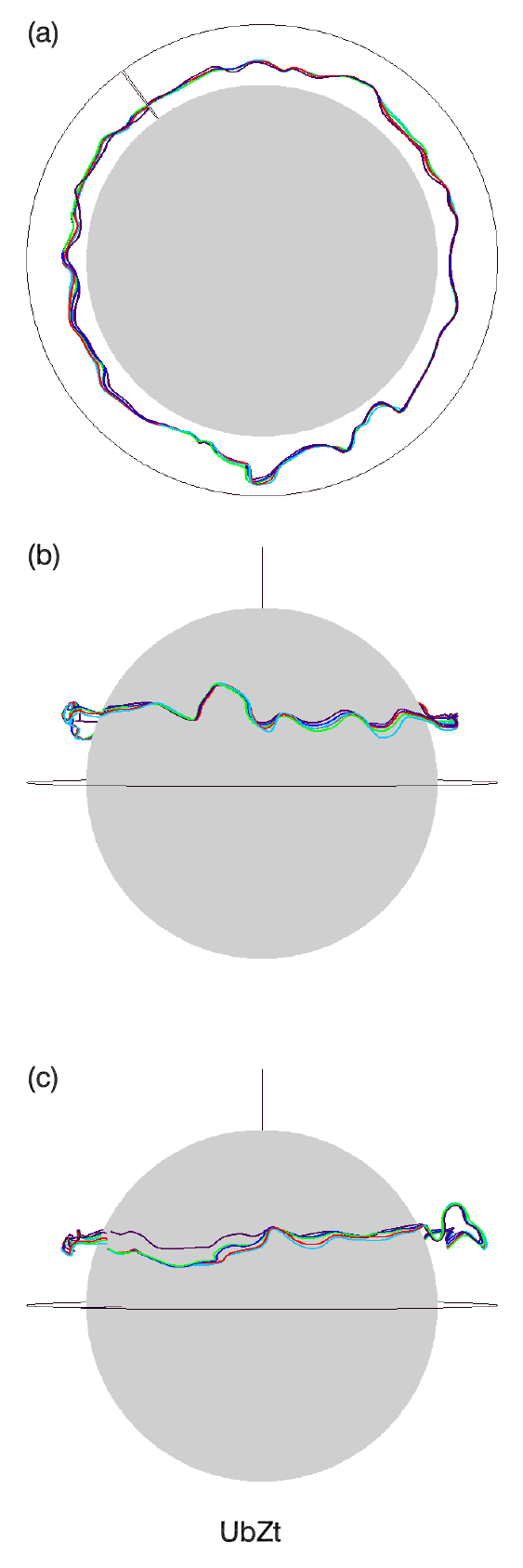}
\caption{Selected field lines in the rising flux tube for the case UbZt as
viewed from the pole with the apex located at the 6 o'clock position (a), and
viewed from the equator with the apex located at the central meridian (b) and
west limb (c) respectively.}
\label{fig:3dfdls_ubzt}
\end{figure}

We have also run a simulation (case UbZt) where the initial toroidal flux
tube is uniformly buoyant along the tube with the magnetic buoyancy, such that
the flux tube would have risen axisymmetrically under its buoyancy had it not
been for the effect of the convective flows.
Thus the development of undulations or loop structures is due entirely to
the convective flows.
Figure \ref{fig:ubzt_3dtube} shows
3D volume rendering of the absolute magnetic field strength of the rising
loops that develop, as viewed from 3 different perspectives, with
the apex portion approaching the top boundary located at the right in
all three views.
An MPEG movie of the evolution of the rising flux tube viewed from the same
perspectives is available in the online version.
We see that loops with shorter footpoint separations form compared to
the 4 major loops formed in the SbZt case.
A set of selected field lines traced from the
apex portion approaching the top are also shown in
a polar view (Figure \ref{fig:3dfdls_ubzt}a) with the apex positioned
at the 6 o'clock location, and two equatorial views 
(Figures \ref{fig:3dfdls_ubzt}b and \ref{fig:3dfdls_ubzt}c) with the
apex positioned at the central meridian and at the west limb respectively.
We can see that despite the fact that the initial buoyancy is uniform
along the tube, the convective
downdrafts are able to hold back portions of the buoyant tube and lead to
the formation of loop structures with undulations that span up to
70\% of the depth of the convection zone (based on the apex and troughs of
the field lines) in a time scale of about a month.
However, the troughs of the loops are not as deeply rooted as the major
loops formed in the SbZt case (compare Figure \ref{fig:3dfdls_ubzt}a with
the top panel of Figure \ref{fig:3dfdls}b).
All of the troughs are above the initial
depth of the toroidal tube, meaning that the downdrafts are not able to
completely overcome the magnetic buoyancy.
Similar to the SbZt case, the convective flows have driven a significantly
larger clockwise tilt from the east-west direction at the apex of the emerging loop, as can be seen in the field line orientation at the central meridian in
Figure \ref{fig:3dfdls_ubzt}b.
of the emerging loop

\begin{figure}
\centering
\includegraphics[width=1.\linewidth]{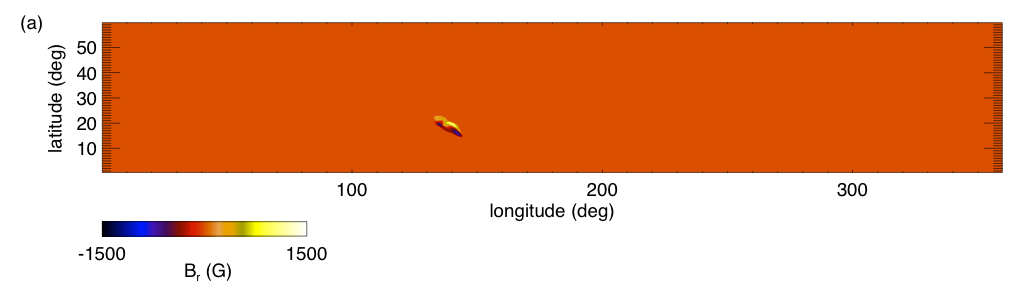}
\includegraphics[width=1.\linewidth]{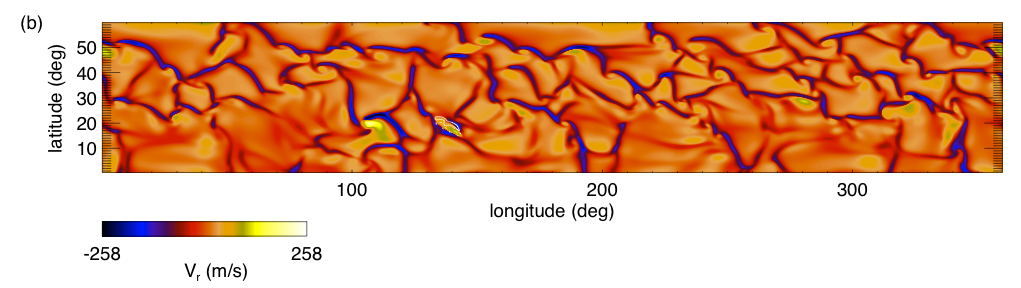}
\includegraphics[width=1.\linewidth]{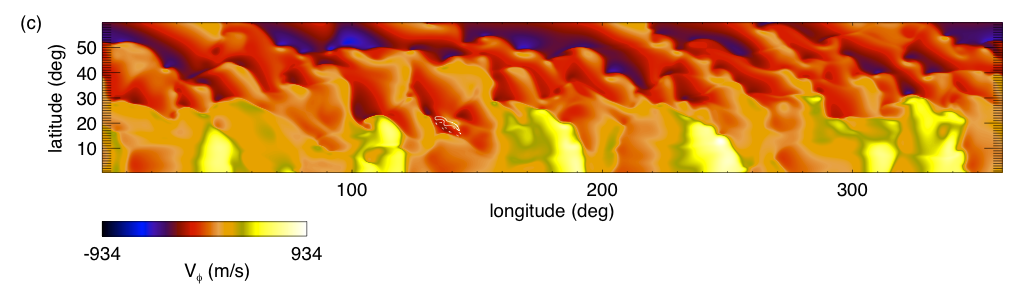}
\caption{Constant $r$ slices of $B_r$ (a), $v_{r}$ (b) and $v_{\phi}$ (c) at
$r=0.957 R_s$ (about 30 Mm below the top boundary)
at the time when the the apex portion of the rising tube is reaching the
top boundary for the UbZt case. White contours in (b) and (c) are contours
of $B_r$ outlining the positive (solid line) and negative (dashed line)
magnetic flux concentrations.}
\label{fig:carrmaps_ubzt}
\end{figure}

Figure \ref{fig:carrmaps_ubzt} shows the normal flux distribution $B_r$,
radial velocity $v_r$, and zonal flow $v_{\phi}$ on a constant $r$ slice
at $r=0.957 R_s$, about 30 Mm below the top boundary, at the time when
the apex portion of the rising flux tube approaches the top boundary for the
UbZt case. An emerging region with a large overall tilt
($75.4^{\circ}$ clockwise based on the flux weighted positions of the
leading and following polarity flux concentrations) of the correct sign
has formed by the apex of the rising tube.
The region of emerging flux corresponds to a local region of upflow (with
speed reaching about 100 m/s) surrounded by narrow downflow lanes (see Figure
\ref{fig:carrmaps_ubzt}b).
The emerging flux also shows a retrograde zonal flow (peaks at about -200 m/s)
in the midst of the prograde flows of the banana cells (see Figure
\ref{fig:carrmaps_ubzt}c).
It corresponds to the most retrograde portion of plasma at that latitude.
Relative to the mean plasma zonal flow speed at that latitude (about 225 m/s),
the emerging flux region has a relative (flux weighted) mean speed of -348 m/s.
Similar results on the relative speeds of the emerging flux
region are found for the SbWt and SbZt cases.

\section{Discussions\label{sec:conc}}

We have used a finite-difference based spherical anelastic MHD code (FSAM) to
simulate rotating solar convection and the buoyant rise of super-equipartition
field strength flux tubes through the convective envelope in the presence
of the giant-cell convection and the associated mean flows.
We achieved a statistically steady solution of giant-cell convection
with a solar-like differential rotation using a relatively low viscosity
$\nu = 10^{12} \, {\rm cm}^2 \, {\rm s}^{-1}$, but a high value of
thermal diffusion $K = 2 \times 10^{13} \, {\rm cm}^2 \, {\rm s}^{-1}$.
The high thermal diffusion allows most of the solar luminosity to be carried
via thermal conduction, so that the resolved giant-cell convection flow speed
is not too high and the convection remains sufficiently rotationally
constrained to give a solar-like differential rotation with the right amplitude.
Into the giant-cell convection near the bottom of the convective envelope,
we insert toroidal flux tubes of 100 kG field strength
and with different forms of magnetic buoyancy distribution to model their
rise through the convective envelope in the presence of convection.
We simulate the rise of the flux tube with no explicit magnetic diffusion
$\eta$ and a quenching of thermal diffusion $K$ in the flux tube to best
preserve the magnetic buoyancy of the initial flux tube.

The simulations show that with a strong, super-equipartition field strength of
100 kG, magnetic buoyancy dominates the rise but the strong down-flows can
significantly modify the shape of the $\Omega$-shaped emerging loops, and
substantially reduce the rise time for the apex to reach the top boundary. Even
if the initial tube is uniformly buoyant, it is found that convection can
produce loop structures with undulations that extend most
of the depth of the CZ in a time scale of about a month. For the weakly
twisted and (initially) untwisted cases we simulated, the apex portion
rises nearly radially and produces an emerging region
with an overall tilt angle consistent with the active region tilts,
although there is continued and substantial loss of flux during the rise.
Thus it appears that the current simulations suggest that a significant twist
in the toroidal magnetic fields in the bottom of the
convection zone is not required for the emergence of coherent active regions.
We emphasize that the current simulations are in a rather laminar region
with the Reynolds number for the rising tube estimated to be $\sim 10$.
This would limit the formation of small scale structures and improve the
cohesion of the rising flux.
However there is difficulty to significantly reduce the viscosity if one
wants to also self-consistently maintain a solar-like
differential rotation \citep[e.g.][]{Featherstone:Miesch:2012}.
On the other hand, the ubiquitous presence of small scale magnetic fields in
a convective dynamo in the CZ may suppress the development of
small scale flows via the magnetic stresses, effectively increasing the
viscosity \citep{Longcope:etal:2003},  and allow a solar
like differential rotation to be maintained at a substantially lower
fluid viscosity (Fan 2013 in preparation).
Thus the presence of the ambient small scale magnetic field
may effectively improve the cohesion of the strong buoyant flux tubes with
weak twists, which is indicated in the recent convective dynamo simulations
in faster rotating convective envelopes by \citet{Nelson:etal:solphys:2013}.
Clearly 3D convective dynamo simulations in the solar convective envelope that
model both the generation of the dynamo mean field and the self-consistent
formation and rise of active region flux in the midst of small scale fields
are needed to obtain a complete understanding of the solar cycle dynamo and
active region formation.

\appendix

\section{The Numerical Algorthms of FSAM}

In this Appendix we describe how FSAM numerically solves equations
(\ref{eq:momentumsolv1}), (\ref{eq:pressuresolv}), (\ref{eq:entropy1solv}),
and (\ref{eq:induction1}), to advance the dependent variables ${\bf v}$,
$p_1$, $s_1$, and ${\bf B}$.
FSAM uses a staggered spatial discretization as described
in \citet{Stone:Norman:1992:a}, where the vector quantities ${\bf v}$ and
${\bf B}$ are defined on the faces of each finite-volume cell of the grid,
scalar quantities $p_1$, $s_1$, $T_1$, are defined at the center
of each finite-volume cell, and the ${\bf v} \times {\bf B}$ electric field
and the current density $\nabla \times {\bf B}$ in the induction equation
are defined on the cell edges. 

First we define some notations to be used frequently
in the rest of the Appendix.
For the spherical polar coordinate system used by this code, we use
subscript $m=1,2,3$ to denote respectively the $r$, $\theta$, $\phi$
direction or component, i.e. we have
$(x_1, x_2, x_3) = (r, \theta, \phi)$,
$(v_1, v_2, v_3) = (v_r, v_{\theta}, v_{\phi})$,
$(B_1, B_2, B_3) = (B_r, B_{\theta}, B_{\phi})$.
Also we make use of the following coordinate scaling
coefficients defined as: $g_2 = r$, $g_{31} = r$, and $g_{32} = \sin \theta $
(notations used in \citet{Stone:Norman:1992:a}).
Consider in general a row of cells in the $m$-direction ($m=1,2,3$),
whose centers' $x_m$ coordinates are located at $x_{m,i}$,
$i=1,2,3,...$, and whose cell averaged $Q$ values are denoted by $Q_i$.
For evaluating the various fluxes at the cell face located
at $x_{m,i-1/2}$ between the two adjacent cells centered on $x_{m,i-1}$
and $x_{m,i}$, we define
\begin{equation}
\delta_m Q \equiv Q_i - Q_{i-1}
\label{eq:delq}
\end{equation}
to be the simple finite difference
between the two adjacent cells (in the m-direction),
and we will use $Q^L$ and $Q^R$ to denote the
`left' and `right' $Q$ values on the cell face, evaluated
through a certain reconstruction of the $Q$ profile within the
cell to the left and right of the cell face, respectively.
Specifically, the assumed profile $Q(x_m)$ within the cell centered
on $x_{m,i}$ is given by a linear reconstruction with a $minmod$ slope limiter:
\begin{equation}
Q(x_m) = Q_i + s_{m,i} (x_m-x_{m,i}),
\label{eq:cellprof}
\end{equation}
where $s_{m,i}$ is a limited slope (in the m-direction) for the cell given by
\begin{equation}
s_{m,i} = minmod \left ( {Q_{i+1} - Q_{i} \over x_{m,i+1} - x_{m,i}},
{Q_{i} - Q_{i-1} \over x_{m,i} - x_{m,i-1}} \right )
\label{eq:s}
\end{equation}
and the $minmod$ function is defined as
\begin{equation}
minmod (y1,y2) \equiv sgn(y1) \max[0,\min(y1, sgn(y1)y2)].
\label{eq:minmod}
\end{equation}
Thus the right and left values, $Q^R$ and $Q^L$,
for the cell face located at $x_{m,i-1/2}$,
between the two neighboring cells centered at $x_{m, i-1}$ and $x_{m,i}$ are:
\begin{equation}
Q^L = Q_{i-1}+s_{m,i-1}(x_{m,i-1/2} - x_{m,i-1}),
\label{eq:ql}
\end{equation}
\begin{equation}
Q^R = Q_{i}-s_{m,i}(x_{m,i} - x_{m,i-1/2}),
\label{eq:qr}
\end{equation}
and we let
\begin{equation}
\Delta_m Q \equiv Q^R - Q^L,
\label{eq:Delq}
\end{equation}
denoting the limited difference between the right and left states at the
cell face at $x_{m,i-1/2}$, and
\begin{equation}
<\!Q\!>_m = {Q^R + Q^L \over 2}
\label{eq:LRmeanq}
\end{equation}
denoting the mean of the left, right values of $Q$ evaluated at the cell
face at $x_{m,i-1/2}$.

The 1,2,3-components of the momentum equation (\ref{eq:momentumsolv1}),
and the entropy equation (\ref{eq:entropy1solv}) we solve written explicitly
in spherical coordinates are:
\begin{eqnarray}
{\partial \over \partial t} ( \rho_0 v_1 )
& = & - {1 \over g_2 g_{31}} {\partial \over \partial x_1}
\left [ g_2 g_{31} \left ( \rho_0 v_1 v_1 \right )^* \right ]
\nonumber \\
& & - {1 \over g_2 g_{32}} {\partial \over \partial x_2}
\left [  g_{32} \left ( \rho_0 v_2  v_1 \right )^* \right ]
\nonumber \\
& & - {1 \over g_{31} g_{32}} {\partial \over \partial x_3}
\left ( \rho_0 v_3 v_1 \right )^*
\nonumber \\
& & + \frac{1}{g_2 g_{32}} \frac{\partial}{\partial x_2} (g_{32} B_1 B_2 )
+ \frac{1}{g_{31} g_{32}} \frac{\partial}{\partial x_3} ( B_1 B_3 )
+ \frac{1}{{g_2}^2 {g_{31}}^2} \frac{\partial}{\partial x_1}
\left ( {g_2}^2 {g_{31}}^2 \frac{{B_1}^2}{2} \right )
\nonumber \\
& & - \frac{1}{{g_{31}}^2} \frac{\partial}{\partial x_1} \left (
\frac{{g_{31}}^2 {B_3}^2}{2} \right ) - \frac{1}{{g_2}^2}
\frac{\partial}{\partial x_1} \left ( \frac{{g_2}^2 {B_2}^2}{2} \right )
\nonumber \\
& & + \rho_0 g_0 \frac{s_1}{c_p}
- \rho_0 {\partial \over \partial x_1} \left ( {p_1 \over \rho_0} \right )
\nonumber \\
& & + \rho_0 v_2^2 {1 \over g_2} {\partial g_2 \over \partial x_1}
+ \rho_0 v_3^2 { 1 \over g_{31}} { \partial g_{31} \over \partial x_1}
+ 2 \Omega \rho_0 v_3 \sin \theta
\nonumber \\
& & + \left [ {1 \over g_2 g_{31}} {\partial \over \partial x_1}
\left ( g_2 g_{31} \rho_0 \nu S_{11} \right ) + {1 \over g_2 g_{32}} {\partial \over \partial
x_2} \left ( g_{32} \rho_0 \nu S_{12} \right ) + {1 \over g_{31} g_{32}} {\partial \over \partial
x_3} \left ( \rho_0 \nu S_{13} \right ) \right ]
\nonumber \\
& & - \rho_0 \nu S_{22} {1 \over g_2}
{\partial g_2 \over \partial x_1} - \rho_0 \nu S_{33}
{1 \over g_{31}} {\partial g_{31} \over \partial x_1}
- {2 \over 3 } {\partial \over \partial x_1}
\left ( \rho_0 \nu \nabla \cdot {\bf v} \right ),
\label{eq:mom1}
\end{eqnarray}
\begin{eqnarray}
{\partial \over \partial t} ( \rho_0 v_2 )
& = & - {1 \over g_2 g_2 g_{31}} {\partial \over \partial x_1}
\left [ g_2 g_{31} g_2^2 \left ( \rho_0 v_1 { v_2 \over g_2 }
\right )^* \right ]
\nonumber \\
& & - {1 \over g_2 g_2 g_{32}} {\partial \over \partial x_2}
\left [  g_{32} g_2^2 \left ( \rho_0 v_2  { v_2 \over g_2 }
\right )^* \right ]
\nonumber \\
& & - {1 \over g_2 g_{31} g_{32}} {\partial \over \partial x_3}
\left [ g_2^2 \left ( \rho_0 v_3 { v_2 \over g_2 } \right )^* \right ]
\nonumber \\
& & + \frac{1}{{g_2}^2 g_{31}} \frac{\partial}{\partial x_1} ( {g_2}^2
g_{31} B_2 B_1 ) + \frac{1}{g_2 {g_{32}}^2} \frac{\partial}{\partial x_2}
\left ( \frac{{g_{32}}^2 {B_2}^2}{2} \right )
\nonumber \\
& & + \frac{1}{g_{31} g_{32}} \frac{\partial}{\partial x_3}
( B_2 B_3 ) - \frac{1}{g_2} \frac{\partial}{\partial x_2}
\left ( \frac{{B_1}^2}{2} \right ) - \frac{1}{{g_{32}}^2 g_2}
\frac{\partial}{\partial x_2} \left ( \frac{{g_{32}}^2 {B_3}^2}{2} \right )
\nonumber \\
& & - {\rho_0 \over g_2} {\partial \over \partial x_2} \left (
{ p_1 \over \rho_0 } \right )
\nonumber \\
& & + \rho_0 v_3^2 {1 \over g_{32} g_2} {\partial g_{32} \over \partial x_2}
+ {2 \Omega } \rho_0 v_3 \cos \theta
\nonumber \\
& & + \left [ {1 \over g_2 g_{31}} {\partial \over \partial x_1}
\left ( g_2 g_{31} \rho_0 \nu S_{21} \right )
+ {1 \over g_2 g_{32}} {\partial \over \partial
x_2} \left ( g_{32} \rho_0 \nu S_{22} \right ) + {1 \over g_{31} g_{32}} {\partial \over \partial
x_3} \left ( \rho_0 \nu S_{23} \right ) \right ]
\nonumber \\
& & + {\rho_0 \nu } S_{21} {1 \over g_2}
{\partial g_2 \over \partial x_1} - {\rho_0 \nu } S_{33}
{1 \over g_{2} g_{32}} {\partial g_{32} \over \partial x_2}
- {2 \over 3 } {1 \over g_2} {\partial \over \partial x_2}
\left ( \rho_0 \nu \nabla \cdot {\bf v} \right ),
\nonumber \\
\label{eq:mom2}
\end{eqnarray}
\begin{eqnarray}
{\partial \over \partial t} ( \rho_0 v_3 )
& = & - {1 \over g_{31} g_{32} g_2 g_{31}} {\partial \over \partial x_1}
\left [ g_2 g_{31} g_{31}^2 g_{32}^2 \left ( \left ( \rho_0 v_1
{ v_3 \over g_{31} g_{32} } \right )^* + \rho_0 v_1 \Omega \right )
\right ]
\nonumber \\
& & - {1 \over g_{31} g_{32} g_2 g_{32}} {\partial \over \partial x_2}
\left [  g_{32} g_{31}^2 g_{32}^2 \left ( \left ( \rho_0 v_2  { v_3 \over
g_{31} g_{32} } \right )^* + \rho_0 v_2 \Omega \right )  \right ]
\nonumber \\
& & - {1 \over g_{31} g_{32} g_{31} g_{32}} {\partial \over \partial x_3}
\left [ g_{31}^2 g_{32}^2 \left ( \left ( \rho_0 v_3
{ v_3 \over g_{31} g_{32} } \right )^* + \rho_0 v_3 \Omega \right ) \right ]
\nonumber \\
& & + \frac{1}{g_{31} g_{32} g_2 g_{31}} \frac{\partial}{\partial x_1}
\left ( g_2 g_{31} g_{31} g_{32} B_3 B_1 \right )
+ \frac{1}{g_{31} g_{32} g_2 g_{32}} \frac{\partial}{\partial x_2}
\left ( g_{32} g_{31} g_{32} B_3 B_2 \right )
\nonumber \\
& & + \frac{1}{g_{31} g_{32}} \frac{\partial}{\partial x_3}
\left ( \frac{{B_3}^2}{2} \right ) - \frac{1}{g_{31} g_{32}}
\frac{\partial}{\partial x_3} \left ( \frac{{B_2}^2}{2} \right )
-\frac{1}{g_{31} g_{32}} \frac{\partial}{\partial x_3} \left (
\frac{{B_1}^2}{2} \right )
\nonumber \\
& & - { \rho_0 \over g_{31} g_{32} } {\partial \over \partial x_3} \left (
{ p_1 \over \rho_0} \right )
\nonumber \\
& & + \left [ {1 \over g_{31} g_{32} g_2 g_{31}} {\partial \over \partial x_1}
\left ( g_2 g_{31} g_{31} g_{32} \rho_0 \nu S_{31} \right )
+ {1 \over g_{31} g_{32} g_2 g_{32}} {\partial \over \partial
x_2} \left ( g_{32} g_{31} g_{32} \rho_0 \nu S_{32} \right ) \right .
\nonumber \\
& &  \left . + {1 \over g_{31} g_{32}} {\partial \over \partial
x_3} \left ( \rho_0 \nu S_{33} \right ) \right ]
- {2 \over 3} {1 \over g_{31} g_{32} } {\partial \over \partial x_3}
\left ( \rho_0 \nu \nabla \cdot {\bf v} \right ),
\label{eq:mom3}
\end{eqnarray}
\begin{eqnarray}
\rho_0 T_0 {\partial s_1 \over \partial t}
& = & - {1 \over g_2 g_{31}} {\partial \over \partial x_1}
\left [ g_2 g_{31} \left ( \rho_0 T_0 v_1 ( s_1+s_0 )
\right )^* \right ]
\nonumber \\
& & - {1 \over g_2 g_{32}} {\partial \over \partial x_2}
\left [  g_{32} \left ( \rho_0 T_0 v_2  ( s_1+s_0 )
\right )^* \right ]
\nonumber \\
& & - {1 \over g_{31} g_{32}} {\partial \over \partial x_3}
\left ( \rho_0 T_0 v_3  ( s_1+s_0 )
\right )^*
\nonumber \\
& & - \rho_0 v_1 (s_1+s_0) {g_0 \over c_p}
\nonumber \\
& & + \rho_0 \nu (S_{12}^2 + S_{23}^2 + S_{31}^2) + {1 \over 6}
\rho_0 \nu \left [ (S_{11} - S_{22})^2 + (S_{22} - S_{33})^2 + (S_{33} - S_{11})^2
\right ]
\nonumber \\
& & + Q_{\rm num} + \eta ( \nabla \times {\bf B} )^2
+ \nabla \cdot \left ( K \rho_0 T_0 \nabla s_1 \right )
\nonumber \\
& & + \nabla \cdot \left ( {16  \sigma_s  T_0^3 \over 3 \kappa \rho_0}
 \nabla T_0 \right ), 
\label{eq:entropy}
\end{eqnarray}
where
\begin{equation}
S_{11} = 2 {\partial v_1 \over \partial x_1},
\end{equation}
\begin{equation}
S_{22} = {2 \over g_2} {\partial v_2 \over \partial x_2} + {2 v_1 \over g_2}
{ \partial g_2 \over \partial x_1}
\end{equation}
\begin{equation}
S_{33} = {2 \over g_{31} g_{32}} {\partial v_3 \over \partial x_3}
+ {2 v_1 \over g_{31}} {\partial g_{31} \over \partial x_1}
+ {2 v_2 \over g_2 g_{32} } {\partial g_{32} \over \partial x_2}
\end{equation}
\begin{equation}
S_{12} = S_{21} = {1 \over g_2} {\partial v_1 \over \partial x_2}
+ g_2 {\partial \over \partial x_1} \left ( {v_2 \over g_2} \right )
\end{equation}
\begin{equation}
S_{23} = S_{32} = {1 \over g_{31} g_{32} } { \partial v_2 \over x_3}
+ {g_{32} \over g_{2}} {\partial \over \partial x_2} \left (
{v_3 \over g_{32} } \right )
\end{equation}
\begin{equation}
S_{31} = S_{13} = {1 \over g_{31} g_{32} } {\partial v_1 \over \partial x_3 }
+ g_{31} {\partial \over \partial x_1 } \left ( {v_3 \over g_{31} } \right ) .
\end{equation}
Note that the 3-component of the momentum equation (\ref{eq:mom3})
is written in the angular momentum conservative form.

For spatially discretizing equations (\ref{eq:mom1}), (\ref{eq:mom2}),
(\ref{eq:mom3}), and (\ref{eq:entropy}), standard 2nd order interpolations
and finite-differences are applied to all the quantities and derivatives,
except for the fluxes (with superscript `*') in the first 3 advection
terms on the right hand side (RHS) of each of the above equations.
For evaluating these fluxes through their respective
cell faces, we use a modified Lax-Friedrichs scheme \citep{Rempel:etal:2009}
to get an upwinded evaluation of the fluxes as follows.

For the first 3 terms on the RHS of equation (\ref{eq:mom1}),
the upwinded evaluation of the 1-, 2-, and 3-fluxes
$\rho_0 v_1 v_1$, $\rho_0 v_2 v_1$, and $\rho_0 v_3 v_1$
through their respective cell-faces at respectively $x_1 = x_{1,i-1/2}$,
$x_2 = x_{2,i-1/2}$, and $x_3 = x_{3,i-1/2}$ are
\begin{equation}
{\left ( \rho_0 v_1 v_1 \right )^* }_{i-1/2} =
\left ( \rho_0 v_1 \right )_{i-1/2} <\!\!v_1\!\! >_1
- \left ( \rho_0 { |v_1|+ c_a \, q_{1,v_1}^l \over 2 } \right )_{i-1/2}
\Delta_1 v_1,
\label{eq:upwindflux_mom1_v1}
\end{equation}
\begin{equation}
{\left ( \rho_0 v_2 v_1 \right )^* }_{i-1/2} =
\left ( \rho_0 v_2 \right )_{i-1/2} <\!\!v_1\!\!>_2
- \left ( \rho_0 { |v_2|+ c_a \, q_{2,v_1}^l \over 2 } \right )_{i-1/2}
\Delta_2 v_1,
\label{eq:upwindflux_mom1_v2}
\end{equation}
\begin{equation}
{\left ( \rho_0 v3 v_1 \right )^* }_{i-1/2} =
\left ( \rho_0 v_3 \right )_{i-1/2} <\!\!v_1\!\!>_3
- \left ( \rho_0 { |v_3|+ c_a \, q_{3,v_1}^l \over 2 } \right )_{i-1/2},
\Delta_3 v_1,
\label{eq:upwindflux_mom1_v3}
\end{equation}
where $<\!\!v_1\!\! >_1$, $<\!\!v_1\!\!>_2$, $<\!\!v_1\!\!>_3$ correspond to
the left-right averages at the respective cell-faces as given by
equation (\ref{eq:LRmeanq}), and $\Delta_1 v_1$, $\Delta_2 v_1$,
$\Delta_3 v_1$ correspond to the limited differences evaluated at the
respective cell-faces as given by equation (\ref{eq:Delq}), and 
\begin{equation}
q_{m,v_1} = {\Delta_m v_1 \over \delta_m v_1},
\label{eq:rempel_factor}
\end{equation}
with $m=1,2,3$, and $\delta_m v_1$ given by equation (\ref{eq:delq}).
Also on the RHS of equations (\ref{eq:upwindflux_mom1_v1}),
(\ref{eq:upwindflux_mom1_v2}), (\ref{eq:upwindflux_mom1_v3}), 
all the other quantities in $()$ are evaluated via standard 2nd
order interpolation at the cell-faces, and $c_a$ denotes the Alfv\'en speed
and $l=4$. The 2nd terms on the RHS of equations (\ref{eq:upwindflux_mom1_v1}),
(\ref{eq:upwindflux_mom1_v2}), and (\ref{eq:upwindflux_mom1_v3}) correspond
to a diffusive flux resulting from the upwinded evaluation
\citep{Rempel:etal:2009}.  
It can be seen that the speed $c_a$ in the diffusive flux is scale by
the smoothness factor $q_{m,v_1}$ (given by equation [\ref{eq:rempel_factor}])
to the $l$th power.  It can be shown that the limited difference
$\Delta_m v_1$ is always of the same sign and of a smaller magnitude compared
to the simple finite difference $\delta_m v_1$. The factor
$q_{m,v_1}^l \ll 1$ when the variation of $v_1$ in the $m$-direction is
smooth, and thus reduces the speed in the diffusive flux.

In the same way, for equation (\ref{eq:mom2}),
the upwinded 1-, 2-, and 3-fluxes $\rho_0 v_1 (v_2 / g_2)$,
$\rho_0 v_2 (v_2 / g_2)$, and $\rho_0 v_3 (v_2 / g_2)$
through their respective cell-faces are:
\begin{equation}
{\left ( \rho_0 v_1 {v_2 \over g_2 } \right )^* }_{i-1/2}
= \left ( \rho_0 v_1 \right )_{i-1/2} <\!\!{ v_2 \over g_2 } \!\! >_1
- \left ( \rho_0 { |v_1|+ c_a \, q_{1,v_2}^l \over 2 } \right )_{i-1/2}
\Delta_1 \left ( { v_2 \over g_2} \right )
\label{eq:upwindflux_mom2_v1}
\end{equation}
\begin{equation}
{\left ( \rho_0 v_2 {v_2 \over g_2 } \right )^* }_{i-1/2}
= \left ( \rho_0 v_2 \right )_{i-1/2} <\!\!{ v_2 \over g_2 } \!\! >_2
- \left ( \rho_0 { |v_2|+ c_a \, q_{2,v_2}^l \over 2 } \right )_{i-1/2}
\Delta_2 \left ( { v_2 \over g_2} \right )
\label{eq:upwindflux_mom2_v2}
\end{equation}
\begin{equation}
{\left ( \rho_0 v_3 {v_2 \over g_2 } \right )^* }_{i-1/2}
= \left ( \rho_0 v_3 \right )_{i-1/2} <\!\!{ v_2 \over g_2 } \!\! >_3
- \left ( \rho_0 { |v_3|+ c_a \, q_{3,v_2}^l \over 2 } \right )_{i-1/2}
\Delta_3 \left ( { v_2 \over g_2} \right ),
\label{eq:upwindflux_mom2_v3}
\end{equation}
where
\begin{equation}
q_{m,v_2} = {\Delta_m ( v_2 / g_2 ) \over \delta_m (v_2 / g_2 )},
\end{equation}
for equation (\ref{eq:mom3}),
the upwinded 1-, 2-, and 3-fluxes $\rho_0 v_1 (v_3 / g_{31} g_{32})$,
$\rho_0 v_2 (v_3 /g_{31} g_{32})$, and $\rho_0 v_3 (v_3 / g_{31} g_{32})$
through their respective cell-faces are:
\begin{equation}
{\left ( \rho_0 v_1 {v_3 \over g_{31} g_{32}} \right )^* }_{i-1/2}
= \left ( \rho_0 v_1 \right )_{i-1/2} <\!\!{ v_3 \over g_{31} g_{32} } \!\! >_1
- \left ( \rho_0 { |v_1|+ c_a \, q_{1,v_3}^l \over 2 } \right )_{i-1/2}
\Delta_1 \left ( { v_3 \over g_{31} g_{32}} \right )
\label{eq:upwindflux_mom3_v1}
\end{equation}
\begin{equation}
{\left ( \rho_0 v_2 {v_3 \over g_{31} g_{32}} \right )^* }_{i-1/2}
= \left ( \rho_0 v_2 \right )_{i-1/2} <\!\!{ v_3 \over g_{31} g_{32} } \!\! >_2
- \left ( \rho_0 { |v_2|+ c_a \, q_{2,v_3}^l \over 2 } \right )_{i-1/2}
\Delta_2 \left ( { v_3 \over g_{31} g_{32}} \right )
\label{eq:upwindflux_mom3_v2}
\end{equation}
\begin{eqnarray}
{\left ( \rho_0 v_3 {v_3 \over g_{31} g_{32}} \right )^* }_{i-1/2}
& = & \left ( \rho_0 v_3 \right )_{i-1/2} <\!\!{ v_3 \over g_{31} g_{32} } \!\! >_3
\nonumber \\
& & - \left ( \rho_0 { |v_3|+ c_a \, q_{3,v_3}^l \over 2 } \right )_{i-1/2}
\Delta_3 \left ( { v_3 \over g_{31} g_{32}} \right ), 
\label{eq:upwindflux_mom3_v3}
\end{eqnarray}
where
\begin{equation}
q_{m,v_3} = {\Delta_m ( v_3 / g_{31} g_{32} )
\over \delta_m (v_3 / g_{31} g_{32} )},
\end{equation}
and finally for equation (\ref{eq:entropy}),
the upwinded 1-, 2-, and 3-fluxes $\rho_0 T_0 v_1 (s_1 + s_0 )$,
$\rho_0 T_0 v_2 (s_1 + s_0 )$, and $\rho_0 T_0 v_3 (s_1 + s_0 )$
through their respective cell-faces are:
\begin{eqnarray}
{\left ( \rho_0 T_0 v_1 (s_1 + s_0 ) \right )^* }_{i-1/2}
& = & \left ( \rho_0 T_0 v_1 \right )_{i-1/2} <\!\! s_1 + s_0 \!\! >_1
\nonumber \\
& & - \left ( \rho_0 T_0 { |v_1|+ c_a \, q_{1,s}^l \over 2 } \right )_{i-1/2}
\Delta_1 \left ( s_1 + s_0 \right ), 
\label{eq:upwindflux_s_v1}
\end{eqnarray}
\begin{eqnarray}
{\left ( \rho_0 T_0 v_2 (s_1 + s_0 ) \right )^* }_{i-1/2}
& = & \left ( \rho_0 T_0 v_2 \right )_{i-1/2} <\!\! s_1 + s_0 \!\! >_2
\nonumber \\
& & - \left ( \rho_0 T_0 { |v_2|+ c_a \, q_{2,s}^l \over 2 } \right )_{i-1/2}
\Delta_2 \left ( s_1 + s_0 \right ), 
\label{eq:upwindflux_s_v2}
\end{eqnarray}
\begin{eqnarray}
{\left ( \rho_0 T_0 v_3 (s_1 + s_0 ) \right )^* }_{i-1/2}
& = & \left ( \rho_0 T_0 v_3 \right )_{i-1/2} <\!\! s_1 + s_0 \!\! >_3
\nonumber \\
& & - \left ( \rho_0 T_0 { |v_3|+ c_a \, q_{3,s}^l \over 2 } \right )_{i-1/2}
\Delta_3 \left ( s_1 + s_0 \right ), 
\label{eq:upwindflux_s_v3}
\end{eqnarray}
where
\begin{equation}
q_{m,s} = { \Delta_m (s_1 + s_0 ) \over \delta_m (s_1 + s_0 ) }.
\end{equation}
Furthermore, in the RHS of the entropy equation (\ref{eq:entropy}) we have also
included a numerical heating term $Q_{\rm num}$ that corresponds to the
dissipation of kinetic energy due to
the diffusive fluxes (the 2nd term in the RHS of eqs.
[\ref{eq:upwindflux_mom1_v1}], [\ref{eq:upwindflux_mom1_v2}],
[\ref{eq:upwindflux_mom1_v3}], [\ref{eq:upwindflux_mom2_v1}],
[\ref{eq:upwindflux_mom2_v2}], [\ref{eq:upwindflux_mom2_v3}],
[\ref{eq:upwindflux_mom3_v1}], [\ref{eq:upwindflux_mom3_v2}],
[\ref{eq:upwindflux_mom3_v3}]), by taking
the dot product of the diffusive fluxes with the appropriate velocity
gradients (computed via the standard centered finite difference),
\begin{eqnarray}
Q_{\rm num} & = & \left ( \rho_0 { |v_1|+ c_a \, q_{1,v_1}^l \over 2 } \,
\Delta_1 v_1 \right ) {\partial v_1 \over \partial x_1}
+ \left ( \rho_0 { |v_2|+ c_a \, q_{2,v1}^l \over 2 } \,
\Delta_2 v_1 \right ) {1 \over g_2} {\partial v_1 \over \partial x_2}
\nonumber \\
& & + \left ( \rho_0 { |v_3|+ c_a \, q_{3,v1}^l \over 2 } \,
\Delta_3 v_1 \right ) {1 \over g_{31} g_{32}} {\partial v_1 \over \partial x_3}
+ \left ( \rho_0 { |v_1|+ c_a \, q_{1,v_2}^l \over 2 } g_2^2 \,
\Delta_1 \left ( {v_2 \over g_2 } \right ) \right )  {\partial \over \partial x_1}
\left ( { v_2 \over g_2 } \right )
\nonumber \\
& & + \left ( \rho_0 { |v_2|+ c_a \, q_{2,v_2}^l \over 2 } g_2^2 \,
\Delta_2 \left ( {v_2 \over g_2 } \right ) \right ) {1 \over g_2}
{\partial \over \partial x_2} \left ( { v_2 \over g_2 } \right )
\nonumber \\
& & + \left ( \rho_0 { |v_3|+ c_a \, q_{3,v_2}^l \over 2 } g_2^2 \,
\Delta_3 \left ( { v_2 \over g_2 } \right ) \right ) {1 \over g_{31} g_{32}}
{\partial \over \partial x_3} \left ( { v_2 \over g_2 } \right )
\nonumber \\
& & + \left ( \rho_0 { |v_1|+ c_a \, q_{1,v_3}^l \over 2 } g_{31}^2 g_{32}^2 \,
\Delta_1 \left ( {v_3 \over g_{31} g_{32} } \right ) \right )
{\partial \over \partial x_1} \left ( {v_3 \over g_{31} g_{32} } \right )
\nonumber \\
& & + \left ( \rho_0 { |v_2|+ c_a \, q_{2,v_3}^l \over 2 } g_{31}^2 g_{32}^2 \,
\Delta_2 \left ( {v_3 \over g_{31} g_{32}} \right ) \right ) {1 \over g_2}
{\partial \over \partial x_2} \left ( {v_3 \over g_{31} g_{32} } \right )
\nonumber \\
& & + \left ( \rho_0 { |v_3|+ c_a \, q_{3,v_3}^l \over 2 } g_{31}^2 g_{32}^2 \,
\Delta_3 \left ( {v_3 \over g_{31} g_{32}} \right ) \right ) {1 \over g_{31} g_{32}}
{\partial \over \partial x_3} \left ( {v_3 \over g_{31} g_{32} } \right ) ,
\label{eq:numheating}
\end{eqnarray}
and then interpolating to the cell centers where $s_1$ is defined.

The pressure equation (\ref{eq:pressuresolv}) we solve can be rewritten as:
\begin{eqnarray}
& & {1 \over \rho_0} {\partial \over \partial x_1}
\left ( g_2 g_{31} \rho_0 {\partial P \over \partial x_1}
\right )
+ {1 \over g_{32}} {\partial \over \partial x_2}
\left ( g_{32} {\partial P \over \partial x_2}
\right )
+ {1 \over g_{32}^2} {\partial^2 P \over \partial x_3^2}
\nonumber \\
& = & {g_2^2 \over \rho_0 } \nabla \cdot {\cal F},
\label{eq:pressure}
\end{eqnarray}
where $P \equiv p_1 / \rho_0$. This linear equation is solved as follows.
The 3-direction ($\phi$-direction) is periodic (for a full $2 \pi$
azimuth), so we carry out a Fourier decomposition of $P$ in the
$x_3$-dimension such that:
\begin{equation}
P_k \equiv P |_{x_3 = x_{3,k}}
= \sum_{n=0}^{N-1} {\hat P}_n e^{i 2 \pi f_n x_{3,k}}
\label{eq:fourier_trans}
\end{equation}
where $x_{3,k}$ with $k=1,2,...N$ are the $N$ uniformly spaced grid points
in $x_3 \in [0,2 \pi]$, $f_n = n/2 \pi$ with $n=0,1,... N-1$ denotes the
discrete spatial frequency, and ${\hat P}_n$
denotes the amplitude of the Fourier component with frequency $f_n$.
Then the centered finite difference evaluation of $\partial^2 P /
\partial x_3^2$ gives:
\begin{equation}
\left ( {\partial^2 P \over \partial x_3^2} \right )_k
= \frac{P_{k+1}-2 P_k + P_{k-1}}{(\delta x_3 )^2}
= \sum_{n=0}^{N-1} \left ( { 2 \cos (2 \pi n / N ) - 2 \over
(\delta x_3 )^2 } \right ) {\hat P}_n e^{i 2 \pi f_n x_{3,k}},
\end{equation}
and equation (\ref{eq:pressure}) leads to the following 2D separable linear
equation for the Fourier component ${\hat P_n} (x_1, x_2)$:
\begin{eqnarray}
& & {1 \over \rho_0} {\partial \over \partial x_1}
\left ( g_2 g_{31} \rho_0 {\partial {\hat P}_n \over \partial x_1}
\right )
+ {1 \over g_{32}} {\partial \over \partial x_2}
\left ( g_{32} {\partial {\hat P}_n \over \partial x_2}
\right )
- {1 \over g_{32}^2} \left ( { 2-2 \cos ( 2 \pi n/N) \over (\delta x_3 )^2 }
\right ) {\hat P}_n
\nonumber \\
& = & {\hat R}_n ,
\label{eq:2dlinear}
\end{eqnarray}
where $\delta x_3$ denotes the grid spacing in $x_3$, and ${\hat R}_n$ is
the Fourier transform of the RHS of equation (\ref{eq:pressure}).
Discretizing the above 2D linear equation leads to a block tridiagonal 
system, which is solved using the routine {\tt blktri.f} in the FISHPACK
math library of the National Center for Atmosphereic Research (NCAR),
based on the generalized cyclic reduction scheme developed by P. Swatztrauber  
of NCAR.

For solving the induction equation (\ref{eq:induction1})
we use the constrained transport (CT) scheme on the staggered grid
\citep{Stone:Norman:1992:b} to ensure the divergence free condition for the
magnetic field (eq. [\ref{eq:divB}]) is satisfied to round-off errors.
The CT scheme is used in conjuction with an upwinded evaluation of
both ${\bf v}$ and ${\bf B}$ based on the Alfv\'en wave characteristics
for computing the ${\bf v} \times {\bf B}$ electric field on the cell
edges as described in \citet{Stone:Norman:1992:b}.
The upwinded evaluation of the
electric field would entail numerical dissipation of the magnetic
field, which we did not put back as heating into the entropy equation.  
Thus, this is a cause of loss of conservation of total energy due to numerical
dissipation in the code.
We also evaluate the physical resistive electric field 
$\eta \nabla \times {\bf B}$ in equation (\ref{eq:induction1}) on the
cell edges following the CT scheme, with the derivatives computed using
simple second order finite differences.
The Ohmic heating produced by the physical resistivity is being included
in the entropy equation (\ref{eq:entropy}).

After the RHS of all of the equations (\ref{eq:mom1}),
(\ref{eq:mom2}), (\ref{eq:mom3}), and (\ref{eq:entropy})
are evaluated at the appropriated cell locations as described above,
we advance the equations in time using a simple second-order
predictor-corrector time stepping.
The linear elliptic pressure equation (\ref{eq:pressure}) is solved
at every sub-timestep to obtain $p_1$ needed for advancing 
equations (\ref{eq:mom1}), (\ref{eq:mom2}), and (\ref{eq:mom3}).


\acknowledgements
NCAR is sponsored by the National Science Foundation.
This work is supported by the NASA SHP grant NNX10AB81G and NASA
LWSCSW grant NNX13AG04A to NCAR.
The numerical simulations were carried out on the Pleiades
supercomputer at the NASA Advanced Supercomputing Division under
project GID s1106.

\clearpage

\clearpage

\clearpage

\clearpage

\clearpage

\clearpage

\clearpage

\clearpage

\clearpage

\clearpage

\clearpage

\clearpage

\clearpage

\clearpage

\clearpage

\clearpage

\clearpage

\clearpage

\clearpage

\clearpage
\begin{deluxetable}{cccccccccccc}
\footnotesize
\tablecaption{Simulation Parameters for FSAM/ASH Comparison \label{inputs_table}}
\tablewidth{0pt}
\tablehead{
\colhead{Parameters}
& \colhead{Values}
}
\startdata
r$_i$ & 4.872$\times$10$^{10}$ cm\\
r$_o$ & 6.960$\times$10$^{10}$ cm\\
$\rho_i$ & 0.36 g cm$^{-3}$\\
$N_{\rho}$ & 1\\
n & 1.5 \\
c$_p$ & 3.5$\times$10$^8$ erg g$^{-1}$ K$^{-1}$\\
M$_{int}$ & 1.989$\times$10$^{33}$ g\\
$\Omega$ & 2.7$\times$10$^{-6}$ s$^{-1}$\\
$\nu$ & 7.74$\times$10$^{12}$ cm$^2$ s$^{-1}$\\
$K$ & 7.74$\times$10$^{12}$ cm$^2$ s$^{-1}$\\
$\Delta S$ & 1543.7 erg K$^{-1}$ g$^{-1}$\\
\enddata
\end{deluxetable}

\begin{deluxetable}{cccccccccccc}
\footnotesize
\tablecaption{Summary of the Rising Flux Tube Simulations \label{tubes_table}}
\tablewidth{0pt}
\tablehead{
\colhead{Label\tablenotemark{a}}
& \colhead{Initial buoyancy}
& \colhead{Twist rate $q \,\,\, (a^{-1})$}
& \colhead{With convection?}
}
\startdata
SbWt-ref & Sinusoidal & -0.15 & No \\
SbZt-ref & Sinusoidal & 0.    & No \\
SbWt & Sinusoidal & -0.15 & Yes \\
SbZt & Sinusoidal & 0.    & Yes \\
UbZt & Uniform  & 0.    & Yes \\
\enddata
\tablenotetext{a}{See \S \ref{sec:risetubesetup} for a
detailed description of the runs}
\end{deluxetable}

\end{document}